\documentclass[lettersize,journal]{IEEEtran}
\usepackage{tikz}
\usepackage{svg}

\usetikzlibrary{shapes.geometric, arrows, positioning}
\providecommand{\keywords}[1]
{

  \textbf{\textit{Keywords---}} #1
}

\tikzset{
  box/.style={rectangle, draw=black, text width=6.5cm, minimum height=1.2cm, text centered, align=center, rounded corners=2pt},
  headerbox/.style={rectangle, draw=black, fill=yellow!30, text width=10cm, minimum height=1cm, text centered, font=\bfseries},
  sidebox/.style={rectangle, draw=black, text width=5.5cm, minimum height=1.2cm, text centered, align=center, rounded corners=2pt},
  labelbox/.style={rectangle, draw=blue, fill=blue!10, minimum width=1.5cm, minimum height=2.5cm, rotate=90, font=\bfseries, align=center},
  mainarrow/.style={thick, -stealth, shorten >=1pt, shorten <=1pt},
  sidearrow/.style={thick, -stealth, shorten >=1pt, shorten <=1pt, dashed}
}

\usepackage{amsmath, amssymb, amsfonts}
\usepackage{graphicx}
\usepackage{textcomp}

\usepackage{hyperref}
\usepackage[caption=false,font=normalsize,labelfont=sf,textfont=sf]{subfig}
\usepackage{algorithm}
\usepackage[noend]{algpseudocode}
\usepackage{multirow}
\usepackage{booktabs} 
\usepackage{colortbl} 
\usepackage{xcolor} 
\usepackage{longtable} 
\usepackage{array} 
\definecolor{highlight}{HTML}{A3C87B}
\usepackage[numbers, sort&compress]{natbib}

\hypersetup{
  colorlinks=true,
  linkcolor=blue,
  citecolor=blue,
  urlcolor=blue
}
\begin{document}

\title{AGI Enabled Solutions For IoX Layers Bottlenecks In Cyber-Physical-Social-Thinking Space}

\author{ 
    Amar Khelloufi\textsuperscript{1},
    Huansheng Ning\textsuperscript{2}{*},
    Sahraoui Dhelim\textsuperscript{3},
    Jianguo Ding \textsuperscript{4}
    \thanks{\textsuperscript{1}Shenzhen Institute of Information Technology, Shenzhen, China}
    \thanks{\textsuperscript{2}University of Science and Technology Beijing, Beijing, China.}
    \thanks{\textsuperscript{3} Dublin City University, Dublin, Ireland.}
     \thanks{\textsuperscript{4} Blekinge Institute of Technology, Karlskrona, Sweden}
     \thanks{\textsuperscript{*} Corresponding Author (ninghuansheng@ustb.edu.cn) }
    \thanks{Manuscript received June 24, 2025;}
    
}

\markboth{Preprint Arxiv}%
{Khelloufi \MakeLowercase{\textit{et al.}}: Preparation of Papers for IEEE Transactions and Journals}


\maketitle

\begin{abstract}
The integration of the Internet of Everything (IoX) and Artificial General Intelligence (AGI) has given rise to a transformative paradigm aimed at addressing critical bottlenecks across sensing, network, and application layers in Cyber-Physical-Social Thinking (CPST) ecosystems. In this survey, we provide a systematic and comprehensive review of AGI-enhanced IoX research, focusing on three key components: sensing-layer data management, network-layer protocol optimization, and application-layer decision-making frameworks.
Specifically, this survey explores how AGI can mitigate IoX bottlenecks challenges by leveraging adaptive sensor fusion, edge preprocessing, and selective attention mechanisms at the sensing layer, while resolving network-layer issues such as protocol heterogeneity and dynamic spectrum management, neuro-symbolic reasoning, active inference, and causal reasoning, Furthermore, the survey examines AGI-enabled frameworks for managing identity and relationship explosion.
Key findings suggest that AGI-driven strategies, such as adaptive sensor fusion, edge preprocessing, and semantic modeling, offer novel solutions to sensing-layer data overload, network-layer protocol heterogeneity, and application-layer identity explosion. The survey underscores the importance of cross-layer integration, quantum-enabled communication, and ethical governance frameworks for future AGI-enabled IoX systems. Finally, the survey identifies unresolved challenges, such as computational requirements, scalability, and real-world validation, calling for further research to fully realize AGI's potential in addressing IoX bottlenecks.
we believe AGI-enhanced IoX is emerging as a critical research field at the intersection of interconnected systems and advanced AI. 

\end{abstract}


\section{Introduction}
\IEEEPARstart{T}{he} Internet of Everything (IoX) is a revolutionary concept that builds on and extends the foundational principles of the Internet of Things (IoT)\cite{chatzigiannakis2021internet}. While IoT primarily involves the connection of physical devices, machines, and sensors to the internet, facilitating the exchange of data and enabling automation, IoX takes this interconnection to a new level by incorporating a broader spectrum of entities\cite{shi2021tutorial}. In addition to connecting physical objects and digital systems, IoX integrates social and cognitive dimensions into the network, creating a more dynamic, adaptive, and context-aware ecosystem. This expanded vision allows for interactions not just between devices, but also between humans, social groups, and intelligent systems, all working together to create smarter environments and more efficient processes\cite{ning2013unit}. While IoT has already transformed various sectors by enabling the remote monitoring, control, and optimization of physical objects—from smart thermostats and wearable health devices to industrial sensors and autonomous vehicles, it is inherently limited to the physical and digital realms. In contrast, IoX goes much further by introducing social and cognitive layers of interaction. The social dimension encompasses the behaviors, preferences, and decisions of individuals, communities, and organizations, enabling systems to respond intelligently to human actions and societal trends\cite{dautenhahn2007socially}. 
IoX comprises three core components—Internet of Things (IoT), Internet of People (IoP), and Internet of Thinking (IoTk)—to form dynamic, context-aware ecosystems within the Cyber-Physical-Social-Thinking (CPST) hyperspace. IoT handles physical interactions, collecting and processing data from devices. IoP focuses on social interactions, capturing behaviors, preferences, and decisions of individuals and communities to enable socially responsive systems\cite{shi2021tutorial}. For example, IoP allows smart cities to adapt services like public transportation based on community needs or social trends. IoTk introduces cognitive capabilities, enabling systems to emulate human-like thinking, learning, and decision-making, such as incorporating mental well-being data in healthcare applications. This multidimensional integration of physical (IoT), social (IoP), and cognitive (IoTk) elements, alongside digital systems (cyber), distinguishes IoX from IoT, fostering intelligent environments in domains like smart cities and healthcare. \\
The evolution of the Internet of People (IoP) emphasizes the integration of human and social interactions into IoT frameworks, forming a critical foundation for the social dimension of Cyber-Physical-Social-Thinking (CPST) ecosystems in IoX \cite{pop00044,miranda2015internet}. IoP shifts the IoX paradigm toward a human-centric ecosystem where devices, with smartphones as central hubs, proactively manage user interactions based on sociological profiles and context-aware reasoning. IoP leverages AGI to process multimodal data—such as location, activity, and social cues—enabling personalized, privacy-preserving interactions and human-like social dynamics\cite{samanta2012}. In contrast, the Internet of Thinking (IoK) focuses on the creation, management, and dissemination of collective knowledge across distributed nodes, utilizing AGI for knowledge discovery, semantic integration, and collaborative intelligence. Together, IoP and IoK enhance the social and cognitive dimensions of CPST ecosystems, complementing IoT’s physical layer focus\cite{miranda2015internet, CONTI201851,zhang2022internet}.
The CPST framework underpins IoX by integrating cyber (digital infrastructure), physical (devices), social (human interactions via IoP), and thinking (cognitive processes via IoTk) spaces, creating interconnected ecosystems that respond to diverse inputs\cite{pasandideh2022cyber}. The cognitive dimension focuses on the ability of both humans and machines to make decisions, learn from experience, and adapt to changing conditions in real time. IoX is not merely about automating processes but also about creating environments that understand, respond to, and optimize interactions across diverse systems, physical, digital, social, and cognitive.\\
This integration of multiple dimensions allows IoX systems to be much more flexible and intelligent than traditional IoT solutions\cite{ning2012cyber,ning2015cyber}. For example, IoX could enable a smart city that not only manages traffic flow and energy usage but also adapts its services based on social behaviors, individual preferences, and cognitive inputs from its inhabitants. Similarly, in healthcare, IoX could enable personalized treatments by considering not just patient data from sensors but also social factors like lifestyle choices, community health trends, and cognitive factors such as mental well-being. The integration of social, cognitive, and physical dimensions in IoX sets the stage for the application of advanced AI systems, particularly Artificial General Intelligence (AGI), which enables more adaptive and intelligent decision-making across these interconnected layers\cite{arshi2024intelligence}.

A comprehensive understanding of cyberspace, as a domain encompassing cyber-enabled interactions, is essential for conceptualizing the CPST hyperspace in IoX \cite{pop00025}. These theoretical frameworks establish the basis for integrating physical, digital, and cognitive elements, necessitating the development of advanced AGI methodologies. As IoT transitions into the broader Internet of Everything (IoX), it introduces complex challenges such as scalability, interoperability, and cross-domain integration, which traditional IoT paradigms struggle to address \cite{pop00007}. AGI’s cross-domain reasoning and autonomous adaptation capabilities are pivotal in overcoming these challenges and enabling more adaptive, resilient IoX systems. A comprehensive review of cyberspace and CPST interactions further reinforces the multidimensional nature of IoX, where cyber, physical, social, and thinking elements converge to form intelligent, interconnected ecosystems \cite{pop00048}. This aligns closely with the concepts of AGI as a transformative enabler for CPST environments. Moreover, the vision of a smart world—characterized by interconnected intelligent systems in applications such as smart cities and healthcare—highlights the real-world potential of IoX \cite{pop00048}. In this context, AGI plays a critical role in overcoming layer-specific bottlenecks and facilitating seamless, efficient integration across diverse domains.

\begin{figure*}
\centering
 \noindent \includegraphics*[width=7.00in, height=2.8in]{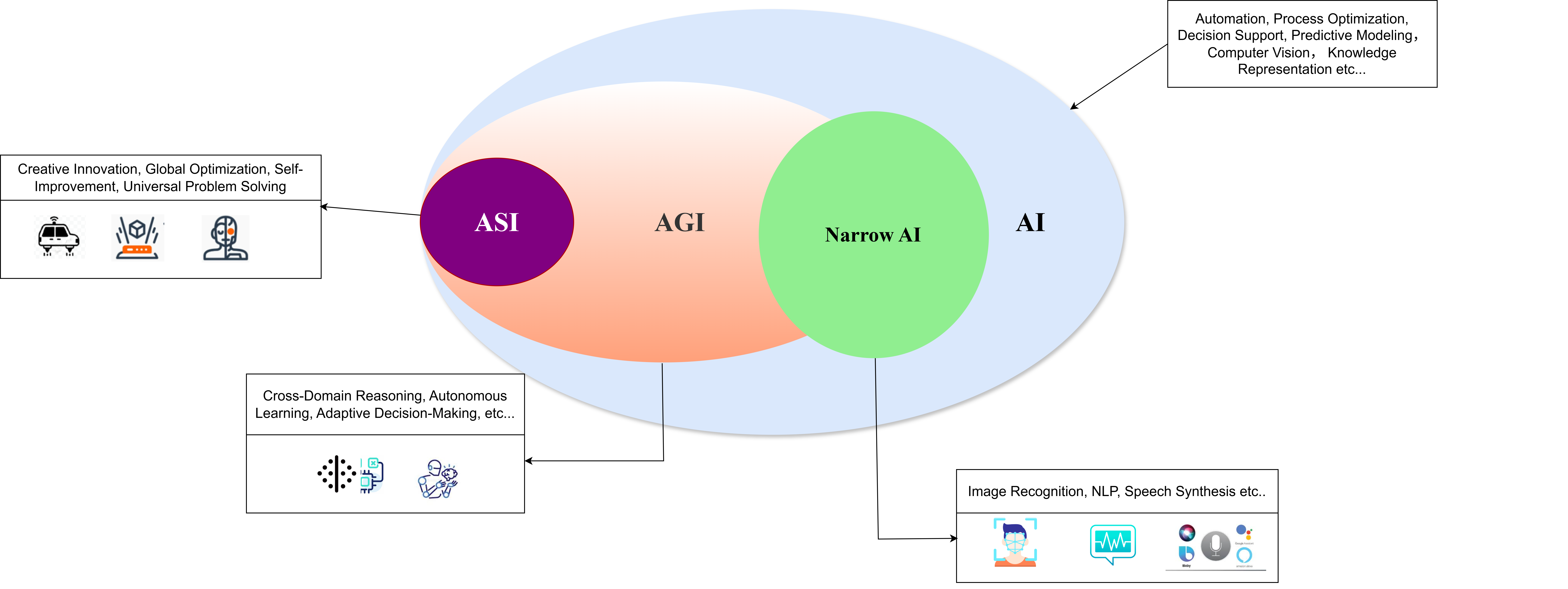}
  \caption{The Hierarchical Relationship of AI, AGI, and ASI}   
\label{AGI_vs_AI}

\end{figure*}

The distinctions between AI, Narrow AI, AGI, and Artificial Superintelligence (ASI) are fundamental to understanding the evolution of intelligent systems\cite{iqbal2024intelligence}. AI encompasses a broad spectrum of technologies designed to perform tasks that typically require human intelligence, ranging from basic automation to complex decision-making. Narrow AI, or also known as weak AI, refers to specialized systems that excel in specific tasks, such as image recognition or language processing, but lack the ability to generalize beyond their predefined functions \cite{babu2024study}.
In contrast, AGI represents a significant advancement in AI, characterized by its ability to perform a wide range of cognitive tasks at a human-like level\cite{morris2023levels}. Unlike Narrow AI, AGI can learn, adapt, and apply knowledge across different domains beyond task-specific programming, enabling more flexible and autonomous decision-making. As such, AGI is considered a crucial milestone in AI research, With the potential to revolutionize multiple fields by enabling more generalized and adaptive intelligence.
Beyond AGI, ASI refers to an intelligence surpassing human capabilities in all cognitive aspects, including creativity, reasoning, and problem-solving\cite{narain2019evolution}. While ASI remains theoretical, AGI is an active area of research, driving advancements toward creating machines with human-like cognitive flexibility. Given its transformative potential, AGI serves as the primary focus of this study, exploring its architectural frameworks, integration challenges, and implications for intelligent systems.
Figure \ref{AGI_vs_AI} depicts the hierarchical relationship among Artificial Intelligence (AI), Narrow AI, Artificial General Intelligence (AGI), and Artificial Superintelligence (ASI), highlighting how each category is encompassed within the broader AI landscape.

Challenges in IoX within CPST
The realization of IoX in Cyber-Physical-Social-Thinking (CPST) ecosystems introduces significant challenges across its layered architecture, which this survey aims to address through AGI-enabled solutions. These challenges, summarized below, arise from integrating IoT, IoP, and IoTk within IoX’s complex, multidimensional framework:
\begin{itemize}
 
\item General Challenges: Scalability, interoperability, and cross-domain integration, driven by the growing number of IoT devices, IoP social systems, and IoTk cognitive processes, necessitating seamless interactions.
\item Sensing Layer Challenges: Data overload and noise from diverse sources, including IoT sensors, IoP social data (e.g., user preferences), and IoTk cognitive inputs (e.g., mental state data), complicating real-time processing and context-aware decision-making.
\item Network Layer Challenges: Scalability, protocol heterogeneity, and bandwidth allocation in heterogeneous environments, exacerbated by IoP’s high-volume social data streams and IoTk’s compute-intensive cognitive processing.
\item Application Layer Challenges: Intelligent decision-making under uncertainty, adapting to dynamic environments, and ensuring security and privacy, particularly for IoP’s socially sensitive data and IoTk’s cognitive models.
\item Multidimensional Integration Challenges: Seamlessly integrating physical (IoT), digital (cyber), social (IoP), and cognitive (IoTk) dimensions to create cohesive ecosystems.
\item Real-World Application Challenges: Adapting to domains like smart cities and healthcare, incorporating IoP-driven social behaviors and IoTk-driven cognitive inputs.

\end{itemize}

\begin{figure*}
\centering
 \noindent \includegraphics*[width=7.00in, height=2.56in]{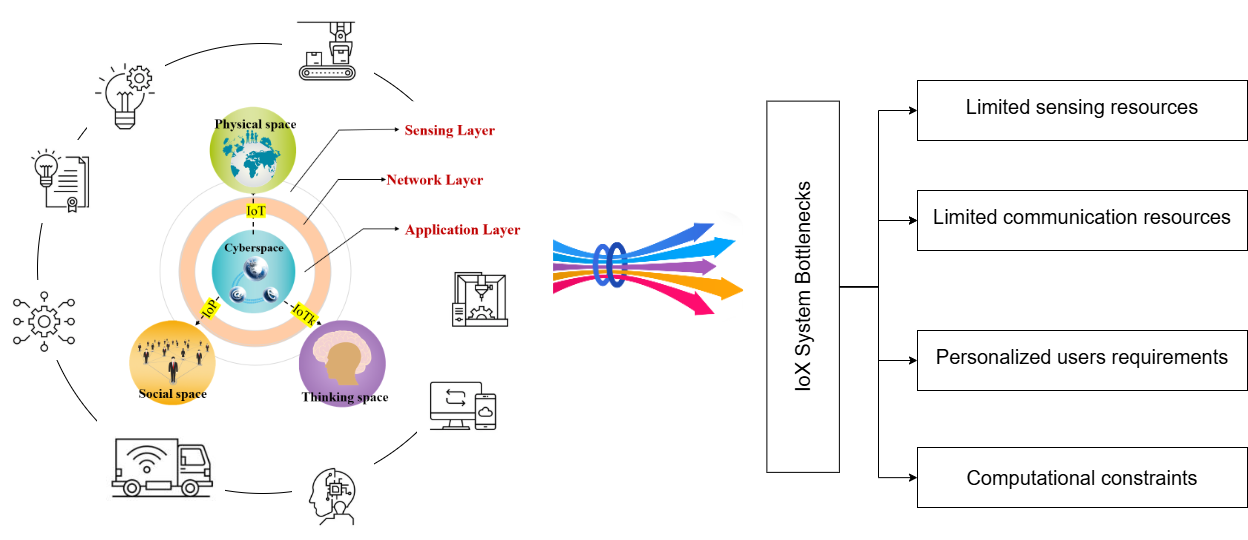}
  \caption{Bottleneck issues rises in IoX}   
\label{Bottleneck_issues}
\end{figure*}

These challenges represent critical bottlenecks that hinder IoX’s potential in CPST ecosystems. This survey focuses on leveraging AGI to mitigate these issues across IoX’s sensing, network, and application layers, as illustrated in Figure \ref{Bottleneck_issues}, which aligns bottlenecks with the CPST model.

The integration of physical, digital, social, and cognitive dimensions in IoX introduces significant challenges across its layered architecture. At the sensing layer, the sheer volume of data generated by interconnected devices often leads to data overload and noise, complicating real-time processing and context-aware decision-making. The network layer struggles with scalability, interoperability, and efficient bandwidth allocation, particularly in heterogeneous environments where devices and protocols must seamlessly interact. Meanwhile, the application layer faces challenges in intelligent decision-making under uncertainty, as well as adapting to dynamic and evolving environments while ensuring robust security and privacy\cite{yao2021security}.

These challenges underscore the need to understand the evolutionary context of IoX and its expanded scope beyond traditional IoT frameworks. The evolution of the Internet of People (IoP) emphasizes the integration of human and social interactions into IoT frameworks, forming a critical foundation for the social dimension of Cyber-Physical-Social-Thinking (CPST) ecosystems in IoX \cite{pop00044}. 
This perspective highlights the need for intelligent systems that adapt to dynamic social behaviors, setting the stage for AGI-driven solutions to address key IoX bottlenecks.

A comprehensive understanding of cyberspace, as a domain encompassing cyber-enabled interactions, is essential for conceptualizing the CPST hyperspace in IoX \cite{pop00025}. These theoretical frameworks establish the basis for integrating physical, digital, and cognitive elements, necessitating the development of advanced AGI methodologies. As IoT transitions into the broader Internet of Everything (IoX), it introduces complex challenges such as scalability, interoperability, and cross-domain integration, which traditional IoT paradigms struggle to address \cite{pop00007}. AGI’s cross-domain reasoning and autonomous adaptation capabilities are pivotal in overcoming these challenges and enabling more adaptive, resilient IoX systems. A comprehensive review of cyberspace and CPST interactions further reinforces the multidimensional nature of IoX, where cyber, physical, social, and thinking elements converge to form intelligent, interconnected ecosystems \cite{pop00048}. This aligns closely with the concepts of AGI as a transformative enabler for CPST environments. Moreover, the vision of a smart world—characterized by interconnected intelligent systems in applications such as smart cities and healthcare—highlights the real-world potential of IoX \cite{pop00048}. In this context, AGI plays a critical role in overcoming layer-specific bottlenecks and facilitating seamless, efficient integration across diverse domains.

These bottlenecks highlight the limitations of traditional approaches, such as Narrow AI and rule-based systems, which often lack the flexibility and adaptability required for IoX’s complex, multi-dimensional ecosystems. AGI, with its ability to generalize knowledge across domains, adapt dynamically, and perform cross-domain reasoning, emerges as a transformative solution. For instance, AGI can enable real-time data fusion and noise reduction in sensor networks, self-organizing and scalable communication protocols, and intelligent, context-aware decision-making that evolves with changing conditions. By addressing these challenges, AGI has the potential to unlock the full potential of IoX in Cyber-Physical-Social-Thinking (CPST) spaces, enabling smarter, more adaptive, and resilient systems.
This survey systematically examines AGI-enabled solutions for overcoming bottlenecks across IoX layers in CPST ecosystems. Our primary focus is on:
\begin {itemize}
\item Architectural Frameworks: How AGI can be integrated into sensing, network, and application layers to enhance performance and adaptability.

\item Integration Challenges: Technical, operational, and scalability barriers in deploying AGI-driven solutions.

\item  Practical Applications: Real-world use cases of AGI in IoX, such as smart cities, healthcare, and industrial automation.
\end{itemize}

Figure \ref{Bottleneck_issues} provides a visual representation of the fundamental bottlenecks in IoX systems, aligned with the CPST (Cyber-Physical-Social-Thinking) model, which organizes the IoX system into four key spaces: Cyber, Physical, Social, and Thinking\cite{zeng2020survey}. The IoX system is composed of three critical layers: the Sensing Layer, which is responsible for collecting raw data from the physical world through various sensors; the Network Layer, which facilitates communication and data transmission between IoX devices and systems; and the Application Layer, where data is processed and transformed into meaningful insights to serve diverse user needs. However, as the number of devices and applications in the IoX ecosystem continues to grow, significant bottlenecks have emerged\cite{cui2021survey}, limiting the performance and scalability of these systems. These challenges include limited sensing resources \cite{perera2013sensor}, which hinder the ability to accurately and comprehensively monitor the environment; limited communication resources\cite{park2016learning}, which result in congestion, data loss, and latency, affecting real-time performance; personalized user requirements\cite{shukla2023improving}, which demand highly customizable and context-sensitive services tailored to individual preferences and needs; and computational constraints, which limit the processing power available for handling the vast amounts of data generated by IoX systems, restricting the implementation of advanced algorithms, AI models, and real-time decision-making processes\cite{ornes2016internet}. 

Effective task allocation among connected devices is critical for addressing computational and energy constraints in IoX, particularly at the sensing and application layers \cite{pop00031}. These challenges highlight the limitations of traditional approaches while underscoring the potential of AGI to optimize resource distribution. The explosive growth of applications, devices, and data links in the IoX ecosystem—emphasized in the figure \ref{Bottleneck_issues} further exacerbates these bottlenecks, making their resolution essential for enhancing system-wide efficiency, adaptability, and sustainability. Ultimately, overcoming these hurdles is pivotal to unlocking IoX’s full potential, where AGI-driven strategies can enable seamless coordination and resilient operation across increasingly complex networks.

While we acknowledge the ethical and societal implications of AGI in IoX, such as privacy concerns and bias in decision-making, these topics are secondary to our technical focus. 
We below highlight the key contributions of the survey, reflecting its comprehensive approach to advancing the understanding and application of AGI in IoX within CPST ecosystems:
\begin{itemize}
    \item Comprehensive Analysis of IoX Challenges: This survey provides a detailed examination of the multidimensional challenges in IoX across sensing, network, and application layers within CPST ecosystems, incorporating IoT, IoP, and IoTk perspectives.
    \item AGI-Driven Solutions: It offers an in-depth exploration of how AGI can address these challenges, highlighting its unique capabilities (cross-domain reasoning, autonomous adaptation, generalized learning) and practical applications in smart cities, healthcare, and industrial automation.
    \item Systematic Literature Review: The study employs a rigorous methodology to synthesize existing research, comparing AGI with Narrow AI and identifying gaps, with a focus on original insights to advance the field.
    \item Architectural and Integration Insights: It proposes frameworks for integrating AGI into IoX layers and discusses technical, operational, and scalability challenges, providing a foundation for future implementations.
    \item Future Research Directions: The survey outlines key areas for future work, including AGI standardization, quantum communication, and application-specific innovations, fostering interdisciplinary collaboration.

\end{itemize}
The rest of this survey is organized as follows:\\
Sections 2 introduces key AGI techniques that underpin solutions to IoX bottlenecks across sensing, network, and application layers. Sections 3 describe the methodology for identifying and analyzing AGI-enhanced solutions to IoX bottlenecks, contrasting AGI’s cognitive capabilities with Narrow AI’s task-specific limitations. Section 4 identifies the IoX system bottlenecks across sensing, network, and application layers, highlighting challenges and improvements through examples like active inference and semantic communications. Section 5 introduces the AGI-driven layer enhancements focusing on implementations, performance, and challenges of AGI-enhanced IoX systems. Section 6 emphasizes the importance of cross-layer integration to optimize AGI-enhanced IoX systems, focusing on coordinated optimization and the technical challenges involved. Section 7 explores future research directions, including testing, AGI standardization, quantum communication, resource management, AI integration, and application-specific innovations like edge intelligence and vehicular network intelligence. The conclusion advocates for interdisciplinary collaboration to overcome technical and regulatory barriers in smart cities, healthcare, and autonomous systems.

\section{AGI Approaches and Methodologies}
To provide a structured understanding of existing methodologies, this section begins with the Core Categories of AGI Architectures in IoX, outlining key architectural approaches based on their foundational principles and relevance to IoT environments.  This is followed by an overview of integrated AGI techniques across the sensing, network, and application layers, illustrating how layered intelligence enables adaptive and context-aware behaviors. We then present a taxonomy of AGI-enhanced IoX architectures, highlighting four dominant paradigms that reflect key design priorities. Finally, we offer a critical analysis of recent studies that exemplify these approaches through practical implementations, drawing insights into their capabilities, challenges, and real-world applicability.

\begin{figure}
  \centering
  \includegraphics[width=3.5in]{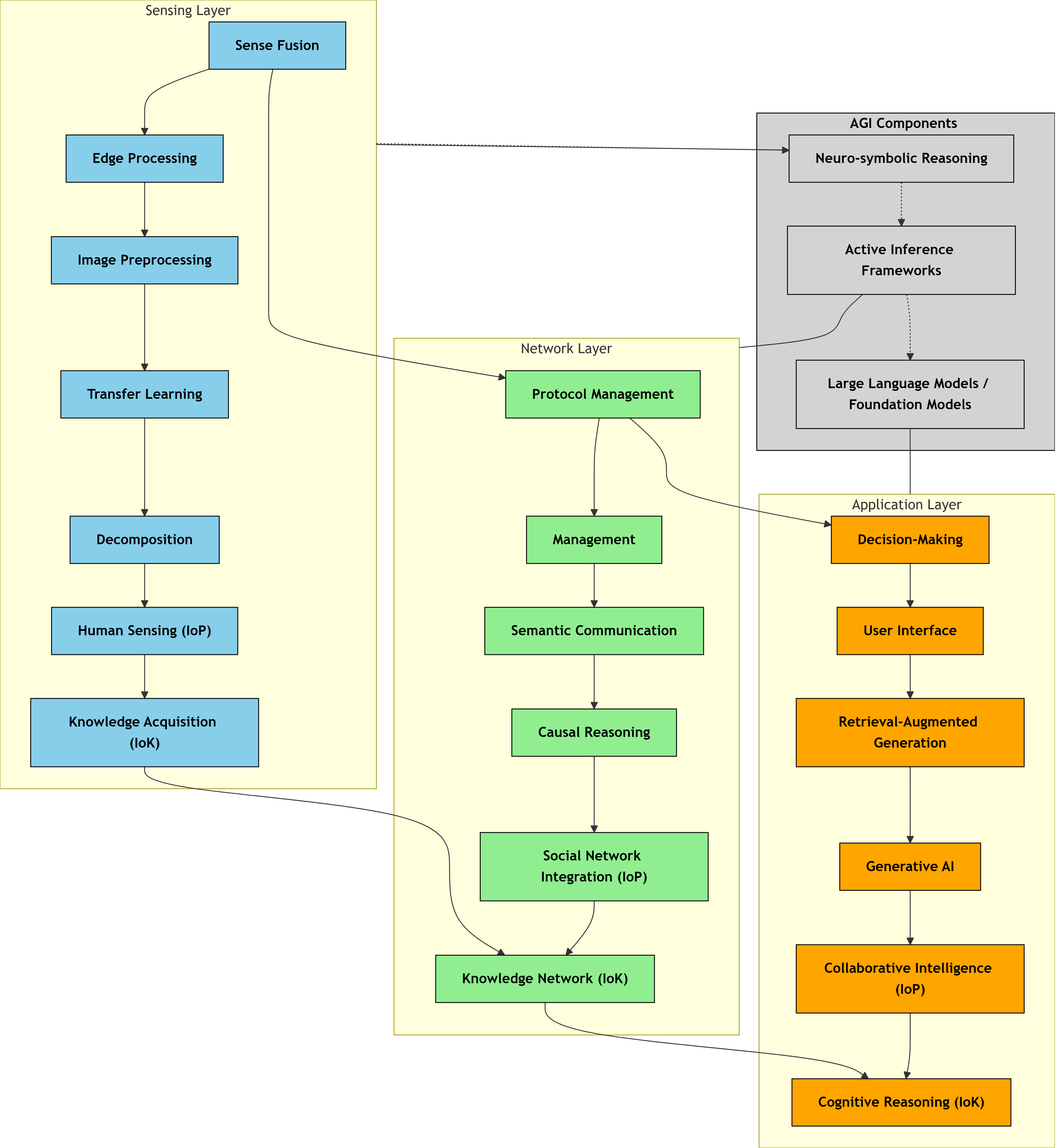}
  \caption{Three-Layer AGI-Enabled IoX Architecture}
  \label{AGI_architecture}
\end{figure}

In order to provide a coherent understanding of the existings methodologies, this section classifies the different AGI architectures into distinct categories based on their underlying principles and their applications within IoT environments. We aim to highlight both their synergies and trade-offs, offering a clearer picture of how each method contributes to the development of smarter, more efficient IoT systems.
\subsection{Core Categories of AGI Architectures in IoX}
The architectural approaches presented here can be broadly grouped into three key categories:
\begin{itemize}
    
\item {Hybrid AI Architectures} This category encompasses approaches that blend different AI techniques to leverage the strengths of both machine learning and symbolic reasoning. Among these, neuro-symbolic AI, Retrieval-Augmented Generation (RAG), and Artificial General Intelligence (AGI) represent innovative strategies that aim to balance performance, robustness, and explainability. These methods are particularly valuable in IoT applications that require sophisticated reasoning and adaptability, while still maintaining transparency and interpretability.

\item Distributed Intelligence: With the increasing scale and complexity of IoT systems, there is a growing need for architectures that distribute computation and decision-making closer to the data source. Edge Intelligence and Federated Learning are two prominent approaches within this theme, both of which emphasize reducing latency, enhancing data privacy, and improving resource efficiency. These architectures are especially relevant in resource-constrained IoT environments, where centralized cloud processing is less feasible or desirable.

\item Generative AI and Language Models: As IoT systems become more context-aware and dynamic, the need for adaptive, content-generating models has increased. Large Language Models (LLMs) such as GPT, and Generative AI (GAI) techniques, are pivotal in enabling IoT systems to generate relevant data, improve decision-making, and enhance overall system responsiveness. These models offer a flexible, data-driven approach to handling the complexity and unpredictability often present in IoT environments.

\end{itemize}
Each of these categories reflects different facets of AI’s role in advancing IoT technology, and in the following sections, we will explore the specific characteristics, strengths, and challenges of each approach. By contextualizing these architectures within real-world use cases and comparing their applications, we will outline how these methods are shaping the future of IoT, from privacy-preserving frameworks to the integration of generative capabilities for enhanced decision-making.

\subsection{Overview of AGI Techniques} 

The integrated AGI architecture represents a hierarchical system where three fundamental layers: sensing, network, and application are stacked vertically, with data flowing upward through the system. At the sensing layer (bottom), raw data collection and preprocessing occur through sensor networks and edge devices, with studies showing unsupervised learning methods for multi-modal feature detection in IoT sensor data \cite{Dmytryk2022Generic}. The network layer (middle) handles data transmission and protocol management, while the application layer (top) focuses on high-level decision-making and user interactions, utilizing hierarchical meta-learning to improve adaptability and precision \cite{Hassan2024Meta}.
These layers are unified through three key AGI components that span across them: Neuro-symbolic reasoning primarily couples the sensing and network layers, enabling improved data efficiency and robustness \cite{Abdelzaher2022Context}. Active inference frameworks bridge network and application layers, facilitating adaptive decision-making and goal-directed behavior \cite{Friston2024Designing}. Large language models and foundation models predominantly operate across application and network layers, enabling sophisticated reasoning and contextual understanding \cite{Guan2024CityGPT}, \cite{Kok2024When}.
Within each layer, specific functionalities are implemented: the sensing layer incorporates multi-modal feature detection and preprocessing \cite{Dmytryk2022Generic}, the network layer manages semantic communications and edge intelligence \cite{Joda2023Internet}, and the application layer handles high-level reasoning and decision-making processes through agent-based foundation models \cite{Xia2024Smart}. This integrated architecture enables comprehensive system optimization through joint sensing, communication, and AI frameworks \cite{Chaccour2024Joint}.\\
Within the IoX architecture, the sensing, network, and application layers serve as the foundational structure supporting three interrelated dimensions: the IoT, IoP, and IoK. AGI techniques enhance these layers by enabling more adaptive, human-centric, and cognitively aware functionalities. Specifically, AGI empowers the application layer to support IoP by facilitating context-aware, proactive interactions using techniques like large language models (LLMs), reinforcement learning, and multi-agent systems. These enable personalized services based on users' sociological profiles, preferences, and behaviors — for example, customizing mobility services in smart cities \cite{miranda2015internet}. Simultaneously, AGI strengthens the network and application layers to support IoK by enabling distributed knowledge discovery, semantic integration, and collaborative intelligence. Generative AI and symbolic reasoning are employed to create real-time, shared knowledge bases across nodes, making it possible to unify heterogeneous data sources for intelligent decision-making in domains such as smart mobility and emergency response \cite{Guan2024CityGPT, Huang2024DriveRP}  By embedding intelligence into these layers, AGI extends the capabilities of IoX to meet the demands of both IoP’s social layer and IoK’s cognitive layer, facilitating a more responsive and knowledge-driven CPST ecosystem. In this way, AGI techniques not only enhance individual IoX layers but also act as enablers for realizing the full potential of IoP and IoK, aligning the system with the multidimensional goals of CPST integration.

Figure \ref{AGI_architecture} illustrates a conceptual AGI-enhanced IoX architecture in which three functional layers—sensing, network, and application—are augmented by cross-layer intelligence modules. The sensing layer at the base is responsible for raw data acquisition and edge-level preprocessing (e.g., unsupervised multi-modal feature detection, brain cognition-inspired frameworks, and active sensing mechanisms). Immediately above, the network layer handles federated learning with blockchain security, information-centric networking, and semantic-native processing, supporting multi-agent systems for distributed resource management. The application layer at the top encompasses mixture-of-experts decision-making and knowledge representation frameworks, leveraging GPT-enhanced reasoning and generative AI services. Bridging these layers are three AGI components: neuro-symbolic explainable AI twins integrate uncertainty-quantified inference across the sensing and network domains; active inference frameworks enable adaptive perception and goal-directed planning between the network and application domains. Together, these elements form a tightly coupled loop of meta-learning and semantic communication protocols, enabling end-to-end optimization of sensing, communication, and reasoning in complex cyber-physical-social-thinking environments.

\section{Research and Survey Methodology}
This section outlines the methodology employed to conduct a comprehensive review of AGI-enhanced solutions for addressing bottlenecks in IoX systems, particularly within Cyber-Physical-Social-Thinking (CPST) frameworks. The methodology encompasses the systematic identification, screening, and analysis of relevant academic literature to address the research question: how AGI can mitigate sensing, network, and application layer challenges in CPST-enabled IoX systems, and what advantages it offers over narrow AI. The approach is structured into three key phases: screening criteria establishment, database collection, and data analysis and synthesis.
\subsection{Screening Criteria}
 To ensure the inclusion of studies most relevant to the research objectives, a rigorous set of screening criteria was developed. These criteria focused on the alignment of studies with AGI applications in IoX or IoT contexts, their relevance to CPST ecosystems, and their contributions to addressing specific layer-based bottlenecks. The screening process involved a holistic evaluation of each study based on the following criteria:

\begin{itemize}

 \item \textbf{AGI Focus:} Studies were required to explicitly address AGI or advanced AI methodologies applicable to IoX or IoT systems. This included approaches such as neuro-symbolic reasoning, active inference, causal reasoning, or foundation models that demonstrate general intelligence capabilities beyond narrow AI.

 \item \textbf{CPST Context:} Studies needed to operate within or be applicable to CPST frameworks, which integrate cyber, physical, social, and cognitive dimensions. This ensured relevance to the complex, interdisciplinary nature of IoX systems.

 \item \textbf{Sensing Layer Relevance:} Studies were evaluated for proposing AGI-driven solutions to sensing layer bottlenecks, such as data overload, energy constraints, noise reduction, or adaptive sampling. Examples include adaptive sensor fusion, edge preprocessing, or selective attention mechanisms.

 \item \textbf{Network Layer Relevance:} Studies were included if they addressed network layer challenges using AGI, such as protocol heterogeneity, dynamic spectrum management, security risks, or intermittent connectivity. Solutions like semantic communications or distributed learning were prioritized.

    \item \textbf{Application Layer Relevance:} Studies tackling application layer issues, such as identity explosion, cognitive feedback loops, or real-time context-aware decision-making (e.g., via semantic modeling or knowledge graphs), were considered for inclusion.

\end{itemize}
Each study was assessed holistically, considering all criteria collectively to determine its relevance. Papers that partially met these criteria but offered significant insights into cross-layer integration or AGI's comparative advantages were also included to ensure a comprehensive review.

\begin{figure*}[t]
\begin{center}
\begin{tikzpicture}[scale=0.70, transform shape, node distance=1.5cm and 1.5cm]
\begin{scope}[xshift=3.4cm]

\node[labelbox, minimum height=1cm, overlay] (id) at (-4.3, -2) {Identification};
\node[labelbox, minimum height=1cm, overlay] (scr) at (-4.3, -7.0) {Screening};
\node[labelbox, minimum height=1cm, overlay] (incl) at (-4.3, -11.5) {Included};

\node[coordinate] (anchor) at (-6.5,0) {};

\node[headerbox, right=2cm of anchor] (title) {Identification of studies via Google Scholar};

\node[box, below=1.0cm of title] (records) 
    {\textbf{Records identified from:}\\ Google Scholar (n = 1320)};
\node[sidebox, right=of records, yshift=-0.0cm] (removed)
    {\textbf{Records removed before screening:}\\ Duplicate records removed (n = 270)\\ Records removed for other reasons (n = 30)};

\node[box, below=1.0 of records] (screened) {Records screened\\(n = 1020)};
\node[sidebox, right=of screened] (excluded1) {Records excluded\textsuperscript{**}\\ (n = 810)};

\node[box, below=1.0 of screened] (retrieved) {Reports sought for retrieval\\(n = 210)};
\node[sidebox, right=of retrieved] (notret) {Reports not retrieved\\(n = 15)};

\node[box, below=1.0 of retrieved] (assessed) {Reports assessed for eligibility\\(n = 195)};
\node[sidebox, right=of assessed] (excluded2) 
    {\textbf{Reports excluded:}\\ Not relevant to AGI (n = 35)\\ Not in CPST context (n = 30)\\ Insufficient detail (n = 32)};

\node[box, below=of assessed] (included) 
    {Studies included in review\\(n = 98)\\ Reports of included studies\\(n = 98)};

\draw[mainarrow] (title) -- (records);
\draw[mainarrow] (records) -- (screened);
\draw[mainarrow] (screened) -- (retrieved);
\draw[mainarrow] (retrieved) -- (assessed);
\draw[mainarrow] (assessed) -- (included);

\draw[sidearrow] (records.east) -- (removed.west);
\draw[sidearrow] (screened.east) -- (excluded1.west);
\draw[sidearrow] (retrieved.east) -- (notret.west);
\draw[sidearrow] (assessed.east) -- (excluded2.west);

\end{scope}
\end{tikzpicture}
\end{center}

\caption{PRISMA flow diagram illustrating the process of study selection for the systematic review
\\ {\small
* Source: Google Scholar only.\\
** Records excluded based on titles and abstracts for not meeting inclusion criteria.
}}
\label{fig:prisma_flow}
\end{figure*}
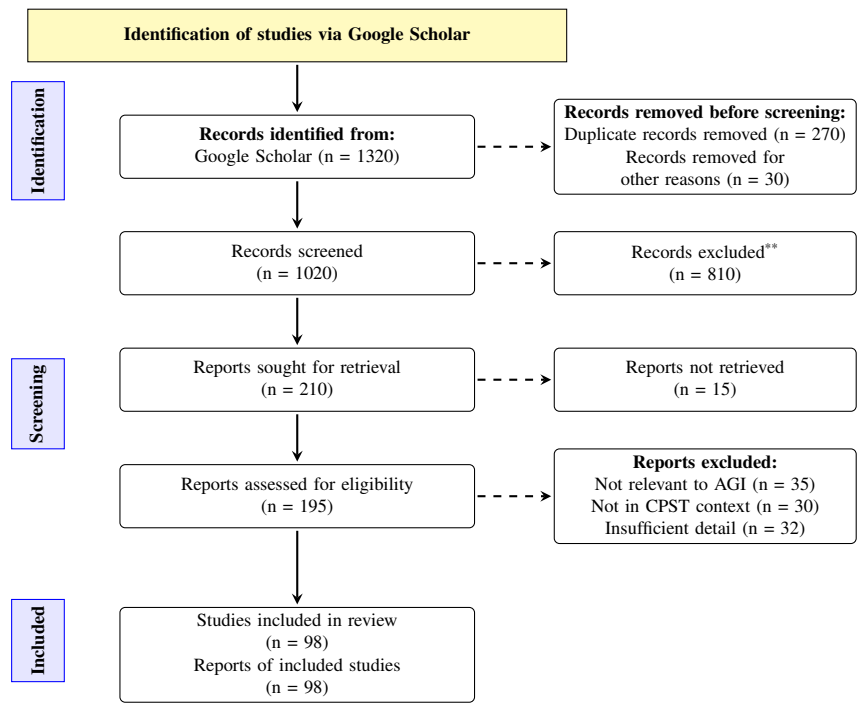

\subsection{Database Collection}
The literature collection process aimed to gather a broad, relevant set of academic papers from authoritative sources. A comprehensive search was conducted using Google Scholar as the primary database due to its extensive indexing coverage across disciplines. The search employed specific search strings to ensure transparency and reproducibility:
\begin{itemize}

\item "Artificial General Intelligence" AND "Internet of " OR "IoX" OR "Internet of everything" 
\item "AGI" AND "Sensing Layer" AND "IoX"
\item "AGI" AND ("Edge Intelligence" OR "Federated Learning") AND "Network Layer"
\item "AGI" AND "Application Layer" AND "CPST"
\item "Internet of People" OR "Internet of Thinking" OR "Internet of Brain"
\end{itemize}
 
The search returned 1320 records. Initial filtering removed 270 duplicates and 30 records for other reasons (e.g., non-English, non-peer-reviewed), leaving 1020 records for screening based on titles and abstracts. Of these, 810 were excluded for not meeting the inclusion criteria. The remaining 210 reports were sought for retrieval, with 195 successfully retrieved. Full eligibility assessment excluded 97 reports for reasons including lack of AGI relevance (n=35), absence of CPST context (n=30), or insufficient detail (n=32). Ultimately, 98 studies were included in the review, matching the PRISMA diagram as depicted in Figure~\ref{fig:prisma_flow}.

To enhance scholarly rigor, key papers were cross-verified using Semantic Scholar, IEEE Xplore, and SpringerLink. This step validated the relevance and quality of included studies and provided full-text access where Google Scholar links were insufficient. These supplementary databases were not used for the initial search but ensured a robust final corpus.

\subsection{Data Analysis and Synthesis}
Following the screening and collection of relevant studies, a systematic approach was employed to analyze and synthesize the data. This phase involved extracting key information from each study, categorizing findings by IoX layer, and identifying thematic insights into AGI's role in addressing CPST-enabled IoX challenges. The analysis was conducted manually to ensure accuracy and depth, focusing on the following aspects:

\begin{itemize}

  \item \textbf{Dimensions of Data Extraction and Analysis:}
A systematic data extraction process was employed to analyze the integration of AGI within IoX systems. This analysis focused on key areas including research methodologies, system architectures, adaptive mechanisms, practical applications, and future research directions. By examining these dimensions, the process provides a comprehensive overview of AGI's role in enhancing IoX ecosystems and its real-world impact across various domains.

\begin{itemize}

\item \textbf{Research Approach and AGI Methodology:} The primary AGI methodology (e.g., cognitive architecture, reasoning framework) was identified, along with key cognitive abilities explored (e.g., perception, reasoning) and novel computational approaches (e.g., hyper-dimensional computing).

\item \textbf{System Architecture and Layer Interactions:} Details on the IoX layers addressed (sensing, network, application) were extracted, with emphasis on AGI mechanisms for cross-layer integration and optimization strategies.

\item \textbf{AGI Adaptive Mechanisms:} Adaptive techniques (e.g., selective attention, edge preprocessing) were documented, noting how they addressed system constraints and incorporated learning or self-optimization.

\item \textbf{Use Case and Practical Applications:} Specific use cases were listed, highlighting AGI-enabled capabilities, performance improvements, and real-world impacts in domains like IoT, IoV, or IoBT.

\item \textbf{Limitations and Future Research:} Explicit limitations of the AGI approaches and recommendations for future research were summarized to contextualize findings and identify gaps.
\end{itemize}

\item \textbf{Thematic Analysis:}

Studies were categorized by their focus on sensing (18 studies), network (25 studies), or application (32 studies) layers, with many addressing multiple layers simultaneously.

Common AGI solutions, such as neuro-symbolic methods and active inference, were identified, along with implementation challenges (e.g., integration issues, resource constraints) and performance impacts (e.g., enhanced decision-making, improved efficiency).

Cross-layer integration strategies, including joint sensing-communication-AI frameworks and federated learning, were analyzed for their technical requirements, benefits, and limitations.

\item \textbf{Comparative Analysis:}

The advantages of AGI over narrow AI were synthesized, emphasizing adaptability, cross-domain reasoning, contextual understanding, explainability, resource efficiency, and holistic optimization.

Implementation considerations, such as edge vs. cloud trade-offs, privacy, scalability, interoperability, real-time performance, energy efficiency, and ethical governance, were highlighted to provide a practical perspective.
\end{itemize}
The synthesis of findings revealed a diverse range of AGI architectures (e.g., neuro-symbolic, LLMs, federated learning) and implementation approaches ( conceptual, experimental, review papers). The analysis underscored the potential of AGI to address IoX bottlenecks while identifying the need for further real-world validation and standardization.

\section{Bottlenecks in IoX Layers}

 \begin{table*}[htbp]
\small
\centering
\caption{IoX Bottlenecks, Affected Layers, and None AGI Mitigation Strategies}
\label{tab:iox-bottlenecks}
\begin{tabular}{@{}>{\centering\arraybackslash}p{4cm}>{\centering\arraybackslash}p{4cm}>{\centering\arraybackslash}p{3cm}>{\centering\arraybackslash}p{5cm}@{}}

\toprule
\textbf{Bottleneck Category} & \textbf{Affected Layers} & \textbf{Impact Severity} & \textbf{None-AGI Proposed Solutions} \\
\midrule

Energy Consumption & Sensing, Network \cite{behzad2018toward} & High & Cross-layer energy optimization, Adaptive protocols \\
\hline

Computational Resources & Application, Sensing \cite{shu2024ris} & Medium & Edge computing, Virtualization \\
\hline

Network Bandwidth & Network \cite{benson2018firedex} & High & SDN-based management, Adaptive routing \\
\hline

Protocol Compatibility & Network, Application \cite{negash2019towards} & Medium & Standardization efforts, Protocol-independent models \\
\hline

Data Exchange & Network, Application \cite{benson2018firedex} & High & Middleware solutions, Cross-domain communication \\
\hline

Cross-layer Communication & All layers \cite{pavkovic2012vers} & Medium & Unified architectures, Adaptive cross-layer designs \\
\hline

Sensing Accuracy & Sensing \cite{duobiene2024enabling} & Medium & Adaptive sensing algorithms, Energy-aware sensing \\
\hline

Network Latency & Network \cite{modekurthy2021low} & High & Low-latency protocols, Edge processing \\
\hline

Processing Capabilities & Application, Sensing \cite{shu2024ris} & Medium & Distributed processing, Optimized resource allocation \\
\hline

\end{tabular}
\end{table*}

The technical bottlenecks in IoX systems form an intricate web of interconnected challenges that span multiple architectural layers, creating systemic inefficiencies that transcend individual components. These impediments manifest differently across sensing, network, and application layers, yet their effects cascade throughout the entire system, producing compound challenges greater than the sum of their individual impacts. At the Sensing Layer, fundamental tensions emerge between data acquisition quality and resource utilization, where issues like sensory overload, environmental noise, and data fusion complexities hinder actionable insight extraction. Limitations in real-time context-awareness further restrict adaptive responsiveness, creating foundational constraints that propagate upward through the system architecture. IoP introduces additional bottlenecks by requiring context-aware processing of multimodal social data (e.g., user preferences), which can exacerbate data overload. AGI-driven selective attention mechanisms, such as those using neuro-symbolic reasoning, mitigate this by prioritizing relevant inputs \cite{miranda2015internet}.

The Network Layer grapples with dual pressures of scalability and interoperability in heterogeneous environments, where bandwidth constraints strain communication efficiency and protocol fragmentation complicates cross-system integration. Simultaneously, the Application Layer contends with decision-making under uncertain or incomplete data streams, while facing persistent security vulnerabilities and the need for dynamic adaptation to evolving user requirements. What makes these layer-specific challenges particularly pernicious is their tendency to interact and amplify across layer boundaries – sensing limitations distort network payloads, network constraints degrade application inputs, and application demands exacerbate resource competition at lower layers.

Understanding these multidimensional bottlenecks requires analyzing both their domain-specific characteristics and their cross-layer implications. The following sections examine these challenges in detail, evaluating their technical origins, systemic impacts, and emerging solutions aimed at enhancing IoX system performance through holistic architectural integration.

\subsection{Network Layer Constraints}

The network layer in IoX systems faces several critical challenges that significantly impact overall system performance, creating a complex ecosystem of interconnected technical bottlenecks that demand careful consideration and management.

\subsubsection{Spectrum Scarcity and Interference}
The primary challenge in this domain relates to the management of limited spectrum resources and interference mitigation. Authors in \cite{bala2022throughput} identify severe interference problems arising from overcrowding in unlicensed frequency bands, which compromises reliable communication and data transmission integrity. This issue is particularly acute in CR-IoT systems, where:
\begin {itemize}
\item Noise uncertainties in low SNR environments lead to deteriorated probability of detection
\item Sensing threshold selection significantly affects overall performance
\item Short packet transmissions require precise optimization of sensing time These factors collectively contribute to reduced system reliability and increased communication failures.
\end{itemize}
\subsubsection{Routing Inefficiencies}
The complexity of routing in IoX networks presents another significant challenge. the study in \cite{kamgueu2017architecture} highlight several critical issues:
\begin{itemize}
\item Multi-hop wireless links lead to suboptimal routing path selection.
\item Coverage area restrictions impact network reach and effectiveness.
\item Path selection algorithms struggle with dynamic network topology changes.
\item Network congestion creates additional routing challenges These routing inefficiencies ultimately affect the system’s ability to efficiently transmit data across the network.
\end {itemize}
\subsubsection{Bandwidth and Latency Issues}
 Bandwidth constraints and their impact on system performance represent a third major challenge. Authors in \cite{modekurthy2021low} demonstrate several key impacts including:
\begin {itemize}
\item Direct relationship between bandwidth constraints and increased latency.
\item Significant impact on base station power dissipation.
\item Quality of Service (QoS) degradation in high-traffic scenarios.
\item Reduced network throughput and increased packet loss rates: The bandwidth limitations create a cascading effect that impacts multiple aspects of network performance. 
\end {itemize}

\subsubsection{Protocol Heterogeneity and Spectrum Scarcity} IoX networks face significant challenges with protocol compatibility and spectrum usage. \cite{negash2019towards} identifies substantial interoperability issues arising from differences in platforms, communication protocols, and data formats. The spectrum environment presents additional challenges, with \cite{bala2022throughput}  highlighting severe interference problems due to overcrowding in unlicensed frequency bands for CR-IoT systems. These protocol and spectrum issues contribute to performance degradation, where \cite{modekurthy2021low} emphasizes the significant impact on latency and base station power dissipation. Integration across different protocols creates additional complications, with \cite{benson2018firedex}  and \cite{bouloukakis2021priodex} identifying network congestion and bandwidth allocation as key challenges in data exchange between heterogeneous IoT components.
Furthermore, the increasing integration of cognitive functions—such as semantic reasoning and distributed knowledge sharing, as envisioned in the IoK exacerbates these challenges. IoK requires semantic interoperability across nodes, adding additional strain on protocol compatibility, bandwidth, and synchronization mechanisms in real-time scenarios\cite{Friston2024Designing}.

\subsection{Sensing Layer Bottlenecks} 
\subsubsection{Energy, Accuracy, and Environmental Constraints}
The interconnected nature of IoX system layers creates a complex web of performance challenges that cascade throughout the entire architecture. At the sensing layer, interference and bandwidth constraints significantly impact data acquisition quality, with authors in \cite{wang2024adaptive} noting that current algorithms struggle to simultaneously optimize sensing accuracy and communication efficiency. These challenges are compounded at the application layer, where authors in \cite{zyrianoff2019architecting} report that latency and routing inefficiencies lead to processing bottlenecks and system crashes, particularly in high-workload scenarios. Authors in \cite{ur2017enhancing} further emphasizes how varying Quality of Service (QoS) requirements across different environments complicate service delivery. \\
Environmental factors pose a significant bottleneck to precise localization, particularly when using RSSI models with BLE beacons. \cite{bravo2021internet} highlight how complex environments, characterized by physical obstacles, signal interference, and mobility constraints, undermine the accuracy of distance estimation. These challenges are further amplified by the need for offline data processing and the limited range of BLE, which requires slower and more deliberate scanning approaches by mobile agents such as UAVs and ground robots\cite{kwe2024emerging}. These constraints reveal critical limitations in achieving seamless and real-time interaction and localization across interconnected IoX systems.

\subsubsection{Performance Management Challenges}
The management of performance challenges in IoX systems demands a sophisticated approach to balancing multiple competing factors and constraints. \cite{khan2020sustainable} demonstrate that effective resource allocation strategies must carefully balance competing demands across different system components, while simultaneously maintaining optimal energy efficiency and latency requirements. Dynamic adaptation mechanisms play a crucial role, with \cite{wang2024adaptive} showing that adaptive sensing algorithms must continuously respond to changing network conditions to maintain system performance. \cite{modekurthy2021low} emphasize the importance of evaluating trade-offs between various performance metrics, noting that improvements in one area often come at the cost of degradation in others, particularly in terms of bandwidth utilization and latency. Authors in \cite{zyrianoff2019architecting} further highlight how system optimization requires careful consideration of both local and global performance impacts, as high workload scenarios can lead to cascading failures across different system components if not properly managed. This complex interplay of factors necessitates a holistic approach to performance management that can dynamically adjust to changing conditions while maintaining optimal system efficiency across all operational parameters.
Building upon these challenges, a critical aspect of maintaining overall IoX system performance lies in effective dataflow management, which faces additional complexities due to data overload and security vulnerabilities, particularly at the sensing and network layers. Dataflow management in IoX systems is hindered by data overload and security vulnerabilities, particularly at the sensing and network layers, which impede real-time processing and reliable communication \cite{pop00001}. AGI’s adaptive data fusion and semantic communication strategies offer promising solutions to these bottlenecks.
These interconnected challenges create a complex web of constraints that must be carefully managed to maintain acceptable network performance levels in IoX systems. The successful implementation of IoX networks requires comprehensive strategies that address these challenges while considering their interdependencies and cross-layer impacts. The complexity of these network layer constraints highlights the need for innovative solutions that can address multiple challenges simultaneously while maintaining system stability and performance. Future developments in IoX systems will need to focus on creating more robust and adaptive networking solutions that can effectively manage these various constraints while supporting the growing demands of IoX applications.

The combined effects of these layer-specific challenges create system-wide bottlenecks that significantly impact overall performance, with \cite{khan2020sustainable} demonstrating that cross-layer optimization approaches can achieve substantial improvements in both latency (51.53\% reduction) and energy consumption (52.88\% improvement). These interconnected challenges underscore the need for holistic solutions that address bottlenecks across all system layers while maintaining optimal performance and reliability.
\subsection{Application Layer Bottlenecks}

The application layer in IoX systems faces critical technical challenges that degrade performance, reliability, and security. Key bottlenecks identified in the literature include:

\subsubsection{High CPU and Memory Demands}
Resource-intensive applications, particularly those deployed in fog or cloud environments, strain computational resources. Studies report that excessive CPU usage and memory limitations in IoT agents lead to system instability and crashes, especially under high workloads \cite{zyrianoff2019architecting}. For instance, general-purpose IoT enablers are often ill-equipped to handle real-time data processing in scenarios like industrial automation or emergency response, resulting in performance degradation.

\subsubsection{Quality of Service (QoS) Variability}
Dynamic QoS requirements across different operational environments (e.g., smart cities vs. vehicular networks) create challenges in resource allocation and prioritization. \cite{ur2017enhancing} highlight that variability in latency, bandwidth, and reliability requirements complicates the design of universally effective IoX applications. This inconsistency is particularly problematic for delay-sensitive use cases, such as autonomous vehicles or healthcare monitoring. 

\subsubsection{Security Vulnerabilities}
The application layer is susceptible to remotely exploitable weaknesses in authentication, access control, and data integrity. For example, authors in \cite{de2024cmxsafe} identify vulnerabilities in service authentication mechanisms that expose systems to cyberattacks. These flaws are exacerbated by the heterogeneous nature of IoX ecosystems, where fragmented security protocols across devices and platforms create entry points for breaches.

\subsubsection{Processing Limitations}
Limited processing capabilities of edge devices constrain real-time decision-making and system responsiveness. In Internet of Robotic Things (IoRT) applications, delays in environmental sensing and computational decision-making hinder mission-critical operations \cite{shu2024ris}. Similarly, \cite{zyrianoff2019architecting} note that high message rates overwhelm application-layer components, leading to bottlenecks in data handling and system crashes. IoP’s focus on social dynamics and IoK’s knowledge management introduce challenges like managing privacy-sensitive social interactions and integrating complex knowledge graphs.
These bottlenecks collectively degrade user experience, compromise system reliability, and expose IoX deployments to security risks. For instance, QoS variability undermines the effectiveness of time-sensitive applications, while processing limitations restrict scalability in large-scale deployments.

Table~\ref{tab:iox-bottlenecks} outlines major IoX bottlenecks, their affected layers, and existing non-AGI solutions. It highlights how challenges at different layers impact overall system performance.

\subsection{Taxonomy of AGI-Enhanced IoX Architectures} 
This section presents a structured taxonomy that organizes AGI-enhanced IoX architectures into a comprehensive, layered framework. Rather than emphasizing abstract architectural paradigms, the taxonomy is grounded in a practical, function-oriented perspective across three foundational layers: Sensing, Network, and Application. Each layer is further divided into key subcategories, reflecting how AGI techniques are applied to address specific technical challenges, including perception, data transmission, decision-making, and coordination.

Figure~\ref{fig:taxonomy} illustrates this taxonomy, which captures the multidimensional integration of AGI within the IoX ecosystem. It delineates how AGI methods—such as neuro-symbolic learning, active inference, generative modeling, and decentralized reasoning—are embedded across architectural layers to enhance system intelligence, efficiency, and autonomy. The sensing layer emphasizes intelligent data acquisition, real-time responsiveness, multi-modal fusion, and autonomous operation. The network layer covers context-aware communication, resource management, edge intelligence, and secure data handling. The application layer focuses on high-level reasoning, urban intelligence, semantic understanding, and multi-agent collaboration.

Beyond the layered view, the taxonomy also considers key cross-cutting dimensions: system architecture , implementation focus , enabling technologies, and application domains. Together, these perspectives provide a coherent and actionable framework to understand how AGI technologies are reshaping the IoX landscape and to guide future innovations in intelligent cyber-physical systems.

\subsubsection{Sensing}
The sensing layer forms the foundational interface between the physical environment and the digital infrastructure of IoX systems. It encompasses mechanisms for data acquisition, fusion, and preprocessing, where AGI technologies play a critical role in elevating system intelligence at the edge. Specifically, AGI enhances perception by integrating neuro-symbolic models, real-time decision-making through active inference frameworks, and multi-modal sensor fusion using adaptive learning. Additionally, this layer increasingly incorporates autonomous capabilities through decentralized agents and reinforcement learning, enabling intelligent sensing strategies that adapt to dynamic environments. These advances support not only high-accuracy data capture but also context-aware interpretation and low-latency responsiveness essential for mission-critical and real-time applications.

\begin{itemize}

\item \textbf{Machine Learning Approaches:}
Intelligent machine learning algorithms are increasingly used to process sensor data efficiently and improve inference in complex environments. Several works propose frameworks that incorporate neuro-symbolic learning for robust, interpretable sensor processing in dynamic scenarios, such as battlefield systems. Others demonstrate how transfer learning and model compression techniques like pruning enable efficient deployment of deep learning models in resource-constrained IoT settings. Additionally, unsupervised clustering methods at the edge help preprocess heterogeneous sensor streams for higher-level cognition in the cloud \cite{Abdelzaher2022Context, Bertino2020Intelligent, Dmytryk2022Generic}. Furthermore, hybrid human-artificial intelligence architectures enhance these frameworks by combining human-like reasoning with computational efficiency, offering robust solutions for complex social and cognitive interactions in IoX systems \cite{pop00005}.

\item \textbf{Real-time Processing:}
Real-time adaptability is achieved through architectures that jointly optimize sensing, communication, and AI processing. Techniques such as tensor decomposition, cognition-inspired decision-making, and edge-based active inference enable fast and adaptive responses in high-interference or mobile environments \cite{Chaccour2024Joint, Ren2024Airground, Devendra2025Towards}. Additionally, the synergy of communication, computing, and caching optimizes smart sensing across all layers, with AGI’s cross-layer integration addressing systemic bottlenecks \cite{pop00043}.

\item \textbf{Multi-modal Sensing:} 
To address the complexity of diverse IoT environments, multi-modal sensing integrates data from various sources using advanced fusion techniques. Research in this area includes meta-learning frameworks that improve predictive accuracy and system efficiency, as well as reconfigurable surfaces that enhance communication fidelity for robotic systems \cite{Hassan2024Meta, Wanli2024Reconfigurable}.

\item \textbf{Autonomous Sensing:}  
Autonomous sensing mechanisms leverage AGI for intelligent decision-making in dynamic environments. This includes multi-agent reinforcement learning for spectrum access in connected vehicles, and decentralized crowdsensing solutions that integrate human and machine contributions, supported by blockchain for trust and verification \cite{Pari2022Secure, Wu2024Autonomous}.
\item \textbf{Context Awareness:}
AGI enhances context awareness by processing multimodal data (e.g., location, activity, social interactions) to create dynamic sociological profiles that capture user preferences and moods. Techniques like neuro-symbolic reasoning and active inference enable real-time adaptation to complex contexts, such as detecting traffic delays or user stress, improving personalization in smart transportation scenarios \cite{miranda2015internet}.

\item \textbf{Proactive Interactions:}
AGI automates device orchestration by reasoning about optimal actions across heterogeneous devices (e.g., smartphones, traffic systems). It enables proactive behaviors, such as rerouting vehicles or suggesting social activities, by leveraging generative AI and reinforcement learning, addressing scalability challenges in large-scale IoP systems \cite{miranda2015internet}.
\end{itemize}

\subsubsection{Network}
The network layer oversees the transmission, routing, and coordination of data across distributed IoX nodes. AGI enhances this layer by introducing intelligent, context-aware communication through techniques such as semantic communications, retrieval-augmented generation (RAG), and AI-assisted protocol adaptation, thereby reducing redundancy and improving data relevance. Additionally, AGI facilitates dynamic resource allocation via reinforcement learning and federated decision-making, optimizing bandwidth, latency, and throughput. Edge intelligence enables real-time inference near data sources, while privacy-preserving mechanisms and blockchain-based verification reinforce security and trust among heterogeneous and autonomous agents.

\begin{itemize}

\item \textbf{Semantic Communications}  
Semantic-aware systems enable more efficient data exchange by interpreting the meaning of transmitted content. This is achieved through AI-driven protocols and natural language processing techniques such as Retrieval-Augmented Generation (RAG), improving communication efficiency and reliability across sensory and vehicular networks \cite{Joda2023Internet, Huang2024DriveRP}. Furthermore, addressing the challenge of exploding connections in IoX relies on advanced AI-driven protocols like semantic communications to ensure efficient data exchange \cite{pop00009}.

\item \textbf{Resource Management:} 
AGI methods optimize network throughput and latency by dynamically managing spectrum and compute resources. Hybrid models employing reinforcement learning and spectrum sharing, along with virtualization at the network edge, support scalable and intelligent resource allocation strategies \cite{VenkataMahesh2024Design, Campolo2021Virtualizing}.

\item \textbf{Edge Computing:}
To reduce latency and enhance responsiveness, distributed intelligence is applied at the edge of the network. Approaches include decentralized learning in Tactile IoT environments and assumed edge AI techniques for real-time adaptation in AGI-enhanced systems \cite{Musa2022Convergence, Do2023Distributed}.

\item \textbf{Security \& Privacy:}
Security solutions combine AGI techniques with federated learning, blockchain, and decentralized decision-making to ensure secure spectrum access, privacy-preserving data sharing, and integrity in integrated space-air-ground networks \cite{Pari2022Secure, Sathwik2025FAPL, Scott2023Secure}.
\end{itemize}
\subsubsection{Application}
The application layer constitutes the cognitive and interactive component of the IoX architecture, encompassing user-facing services, system-level reasoning, and collaborative intelligence. AGI technologies at this layer enable autonomous decision-making, semantic understanding via large language models (LLMs) and knowledge graphs, and adaptive responses in dynamic environments. Urban intelligence platforms utilize AGI to integrate spatiotemporal data and deliver scalable, real-time solutions to societal and environmental challenges. Furthermore, generative AI and hybrid symbolic-neural models enhance knowledge representation and scenario planning, while collaborative multi-agent systems support distributed problem-solving in areas such as mobility, energy, and public safety.

\begin{itemize}

\item \textbf{Autonomous Systems:} 
Advanced driving and smart mobility applications benefit from AGI by incorporating foundation models, generative AI, and scenario-aware decision-making. These systems leverage multi-modal inputs for improved navigation, reasoning, and planning in autonomous settings \cite{Huang2024DriveRP, Xia2024Smart, Xie2024GAI}.

\item \textbf{Urban Intelligence:}
Urban environments are becoming smarter through multi-agent systems powered by large language models. These enable real-time analytics, spatiotemporal data visualization, and secure information sharing during crises \cite{Guan2024CityGPT, Yue2024paradigm}. Moreover, hybrid human-artificial intelligence enhances social computing by enabling IoX systems to model and respond to social dynamics, advocating for AGI’s role in social intelligence within CPST frameworks \cite{pop00064}.

\item \textbf{Knowledge Representation:} 
Knowledge modeling in AGI-enabled IoX systems is supported by structured graph architectures, symbolic AI, and active inference. These methods enable systems to generalize, adapt, and provide meaningful insights in real time across diverse IoT domains \cite{Di2022System, Hugo2022Hybrid, Friston2024Designing}.

\item \textbf{Multi-agent Systems:}
Collaborative intelligence is achieved through generative agents and interaction-aware frameworks. These systems enhance adaptability and reasoning in environments such as smart infrastructure and 6G networks, where communication efficiency and decision coordination are critical \cite{Nascimento2023GPT, Zou2024GenAINet, Xu2024Integration}.
\end{itemize}
Additionally, the exploration of hybrid human-artificial intelligence for multimedia computing enhances context-aware decision-making and multi-agent coordination, supporting AGI integration in CPST ecosystems \cite{pop00086}.
\subsubsection{Human-Centric Intelligence}

Human-centric AGI integration in IoX emphasizes adaptability, memory retention, emotional awareness, and personalized interaction, enabling systems to align more closely with human expectations and behavior.

\begin{itemize}

\item \textbf{Modular Architectures:}
Human-aligned AGI frameworks increasingly adopt modular architectures comprising perception, world modeling, and action-planning components. These architectures leverage hyper-dimensional computing to dynamically allocate network resources in immersive environments such as digital twins and XR applications, supporting seamless connectivity and responsiveness \cite{Saad2024Artificial}.

\item \textbf{Interactive Frameworks:}
Adaptive decision-making in IoX is supported by interactive frameworks that structure intelligence across environment sensing, perception, cognitive reasoning, and actuation. By integrating LLMs and retrieval-augmented generation (RAG), these systems achieve unified understanding across heterogeneous IoP data sources and enable responsive, context-aware network control \cite{Zhang2024Interactive}.

\item \textbf{Context Retention:}
AI-native memory systems underpin personalization by incorporating deep neural memory layers and Lifelong Personal Models. These systems continuously learn from user behavior to retain relevant contextual information, enabling proactive engagement in applications such as memory-augmented dialogue, recommendations, and assistive services \cite{shang2024ai}.

\item \textbf{Social and Emotional Intelligence:}
AGI systems are evolving to recognize and respond to emotional and social cues, fostering empathetic human-machine interaction. This capability is vital in domains such as personalized tutoring, collaborative teamwork, and social robotics, where systems must interpret nuanced human intent and adapt communication strategies accordingly \cite{cichocki2020future, latif2023agi}.

\end{itemize}

\twocolumn

\tikzstyle{role} = [rectangle, rounded corners, minimum width=0.5cm, minimum height=1cm, align=left, draw=black, anchor=west]
\tikzstyle{Mainrole} = [rectangle, rounded corners, minimum width=2cm, minimum height=1cm, align=left, draw=black, rotate=90, fill=blue!20]
\tikzstyle{Mainrole2} = [shape=ellipse, minimum width=2cm, minimum height=1cm, align=left, draw=black, rotate=90, fill=blue!20]

\tikzstyle{role1} = [rectangle, rounded corners, minimum width=0.5cm, minimum height=0.5cm, align=left, draw=black, anchor=west]

\begin{figure*}[ht]
    \centering
\begin{tikzpicture}
        \node (A) [Mainrole] at (0,0) {Taxonomy};
        \node (F) [Mainrole] at (1.5,-7) {Application};
        \node (C) [Mainrole] at (1.5,0) {Network};
        \node (B) [Mainrole] at (1.5,6) {Sensing};
        \node (D) [Mainrole2, align=center] at (1.5,-11) {IoP \& IoK \\ Human Centric};

        \node (B1) [role, align=center] at (6.5,8.5) {Machine Learning \\Approaches };
      
        \node (B2) [role, align=center] at (6.5,6.5) {Real-time Processing};
        \node (B3) [role, align=center] at (6.5,4.8) {Multi-modal Sensing};
        \node (B4) [role, align=center] at (6.5,3.4) {Autonomous Sensing};
        
        \node (B5) [role, align=center] at (6.5,1.9) {Semantic\\ Communications};
        \node (B6) [role, align=center] at (6.5,0.7) {Resource\\ Management};
        \node (B7) [role, align=center] at (6.5,-0.5) {Edge Computing};
        \node (B8) [role, align=center] at (6.5,-2) {Security \& Privacy};
        
        \node (B9) [role, align=center] at (6.5,-4) { Autonomous Systems};
        \node (B10) [role, align=center] at (6.5,-5.7) {Urban Intelligence};
        \node (B11) [role, align=center] at (6.5,-7.2) {Knowledge\\ Representation};
        \node (B12) [role, align=center] at (6.5,-8.8) {Multi-agent Systems};

        \node (F1) [role1] at (10.5,8.5) {Neuro-symbolic ML \\ Transfer learning/pruning \\ Unsupervised learning };
        \node (F2) [role1] at (10.5,6.5) { Tensor decomposition \\ Brain cognition-inspired frameworks \\ Active sensing};
        \node (F3) [role1] at (10.5,4.8) {Meta-learning frameworks\\ Reconfigurable surfaces};
        \node (F4) [role1] at (10.5,3.4) {Spectrum sensing\\ Autonomous crowdsensing};
        \node (F5) [role1] at (10.5,1.9) {Semantic frameworks\\ RAG-based AI  };
        \node (F6) [role1] at (10.5,0.7) {Collaborative management \\Network virtualization};
        \node (F7) [role1] at (10.5,-0.5) {Edge intelligence\\ Distributed ML};
        \node (F8) [role1] at (10.5,-2) {Secure spectrum access \\Federated learning \\ Secure SAGIN };
         \node (F9) [role1] at (10.5,-4) { RAG for driving \\ Mobility foundation models \\ GAI-IoV frameworks};
        \node (F10) [role1] at (10.5,-5.7) {CityGPT \\ Fusion networks};
        \node (F11) [role1] at (10.5,-7.2) {System-level knowledge\\  Hybrid AI \\Active inference};
        \node (F12) [role1] at (10.5,-8.8) {GPT-in-the-loop \\GenAINet \\ MoE/GAI integration};
        \node (J1) [role1] at (15,8.5) {\cite{Abdelzaher2022Context,Bertino2020Intelligent,Dmytryk2022Generic} };
        \node (J2) [role1] at (16.5,6.5) {\cite{Chaccour2024Joint,Ren2024Airground, Devendra2025Towards}};
        \node (J3) [role1] at (15,4.8) {\cite{ Hassan2024Meta, Wanli2024Reconfigurable}};
        \node (J4) [role1] at (15,3.4) {\cite{Pari2022Secure,  Wu2024Autonomous}};
        \node (J5) [role1] at (15,1.9) {\cite{Joda2023Internet, Huang2024DriveRP}};
        \node (J6) [role1] at (15,0.7) {\cite{VenkataMahesh2024Design, Campolo2021Virtualizing}};
        \node (J7) [role1] at (15,-0.5) {\cite{Musa2022Convergence, Do2023Distributed}};
        \node (J8) [role1] at (15,-2) { \cite{Pari2022Secure, Sathwik2025FAPL, Scott2023Secure}};
         \node (J9) [role1] at (15.5,-4) {\cite{Huang2024DriveRP, Xia2024Smart, Xie2024GAI}};
        \node (J10) [role1] at (15,-5.7) {\cite{Guan2024CityGPT, Yue2024paradigm}};
        \node (J11) [role1] at (15,-7.2) {\cite{Di2022System, Hugo2022Hybrid, Friston2024Designing}};
        \node (J12) [role1] at (15,-8.8) {\cite{Nascimento2023GPT, Zou2024GenAINet, Xu2024Integration}};

        
        \node (K1) [role, align=left ] at (3,7.8) {Data Processing\\ \& Feature \\Extraction};
        \node (K2) [role, align=left ] at (3,4) {Sensor Integration\\ \& Management};
        \node (K3) [role, align=left ] at (3,1.4) {Communication \\Protocols};
        \node (K4) [role, align=left ] at (3,-1.1) {Intelligence \\Distribution};
        \node (K5) [role, align=left ] at (3,-4.8) {Decision \\Support Systems};
        \node (K6) [role, align=left ] at (3,-8) {Cognitive \\Services};
       
        \node (D1) [role, align=left ] at (3,-10.5) {Human Dynamics};
        \node (D2) [role, align=left ] at (3,-12.2) {Social Networks};
        \node (D3) [role, align=left ] at (6.5,-10.5) {Human Behavior \\ Modelling};
        \node (D4) [role, align=left ] at (6.5,-12.2) {Social Network\\ Analysis};
        \node (D5) [role, align=left ] at (10.5,-10.5) {Cognitive digital twins \\Social physics models\\ Behavior prediction };
        \node (D6) [role, align=left ] at (10.5,-12.2) {Network Centrality \\Community Structure\\Information Cascades \\ Proactive Engagement};
        \node (D7) [role1, align=left ] at (15,-10.5) {\cite{nechesov2025virtual,CONTI201851, abbasi2015predicting,cichocki2020future, latif2023agi} };
        \node (D8) [role1, align=left ] at (15,-12.2) {\cite{galik2020thinking, breslin2007future, zhang2022internet, shang2024ai} };
        \draw (A) edge (0.7,0);
        \draw (0.7,-11) edge (0.7,6);
         \draw (0.7,-11) edge (D);
        \draw (0.7,6) edge (B);
        \draw (0.7,0) edge (C);
        \draw (0.7,-7) edge(F);
        \draw (2.5,7.8) edge (2.5,4);
        \draw (2.5,7.8) edge (K1);
        \draw (2.5,4) edge (K2);
        \draw (2.5,1.4) edge (2.5,-1.1);
        \draw (2.5,1.4) edge (K3);
        \draw (2.5,-1.1) edge (K4);
        \draw (2.5,-4.8) edge (2.5,-8);
        \draw (2.5,-4.8) edge (K5);
        \draw (2.5,-8) edge (K6);
        \draw (D) edge (2.5,-11);
        \draw (2.5,-10.5)edge (2.5,-12.2);
        \draw(2.5,-10.5)edge (D1);
        \draw(2.5,-12.2)edge(D2);
        \draw(D1)edge(D3);
        \draw(D2) edge(D4);
        \draw(D3) edge(D5);
        \draw(D4) edge(D6);
         \draw(D5) edge(D7);
          \draw(D6) edge(D8);
        \draw (B) edge (2.5,6);
        \draw (C) edge (2.5,0);
        \draw (F) edge (2.5,-7);
        \draw (6.3,8.5) edge (6.3,6.5);
        \draw (6.3,4.8) edge (6.3,3.4);
        \draw (6.3,1.9) edge (6.3,0.7);
        \draw (6.3,-0.5) edge (6.3,-2);
        \draw (6.3,-4) edge (6.3,-5.7);
        \draw (6.3,-7.2) edge (6.3,-8.8);
        \draw (K1) edge (6.3,7.8);
        \draw (K2) edge (6.3,4);
        \draw (K3) edge (6.3,1.4);
        \draw (K4)edge (6.3,-1.1);
        \draw (K5)edge (6.3,-4.8);
        \draw (K6)edge (6.3,-8);
        \draw (6.3,8.5) edge (B1);
        \draw (6.3,6.5) edge (B2);
        \draw (6.3,4.8) edge (B3);
        \draw (6.3,3.4) edge (B4);
        \draw (6.3,1.9) edge (B5);
        \draw (6.3,0.7) edge (B6);
        \draw (6.3,-0.5) edge (B7);
        \draw (6.3,-2) edge (B8);
        \draw (6.3,-4) edge (B9);
        \draw (6.3,-5.7) edge (B10);
        \draw (6.3,-7.2) edge (B11);
        \draw (6.3,-8.8) edge (B12);
        \draw (B1) edge (F1);
        \draw (B2) edge (F2);
        \draw (B3) edge (F3);
        \draw (B4)edge (F4);
        \draw (B5)edge (F5);
        \draw (B6)edge (F6);
        \draw (B7)edge (F7);
        \draw (B8)edge (F8);
        \draw (B9)edge (F9);
        \draw (B10)edge (F10);
        \draw (B11)edge (F11);
        \draw (B12)edge (F12);
        \draw  (F1) edge (J1);
        \draw  (F2)edge (J2);
        \draw  (F3)edge (J3);
        \draw  (F4)edge (J4);
        \draw  (F5)edge (J5);
        \draw  (F6)edge (J6);
        \draw  (F7)edge (J7);
        \draw  (F8)edge (J8);
        \draw  (F9)edge (J9);
        \draw  (F10)edge (J10);
        \draw  (F11)edge (J11);
        \draw  (F12)edge (J12);


\end{tikzpicture}
    \caption{Taxonomy of AGI-enabled solutions for IoX layers (IoT, IoP, IoK) in cyber-physical-social-thinking space.}
    \label{fig:taxonomy}
\end{figure*}
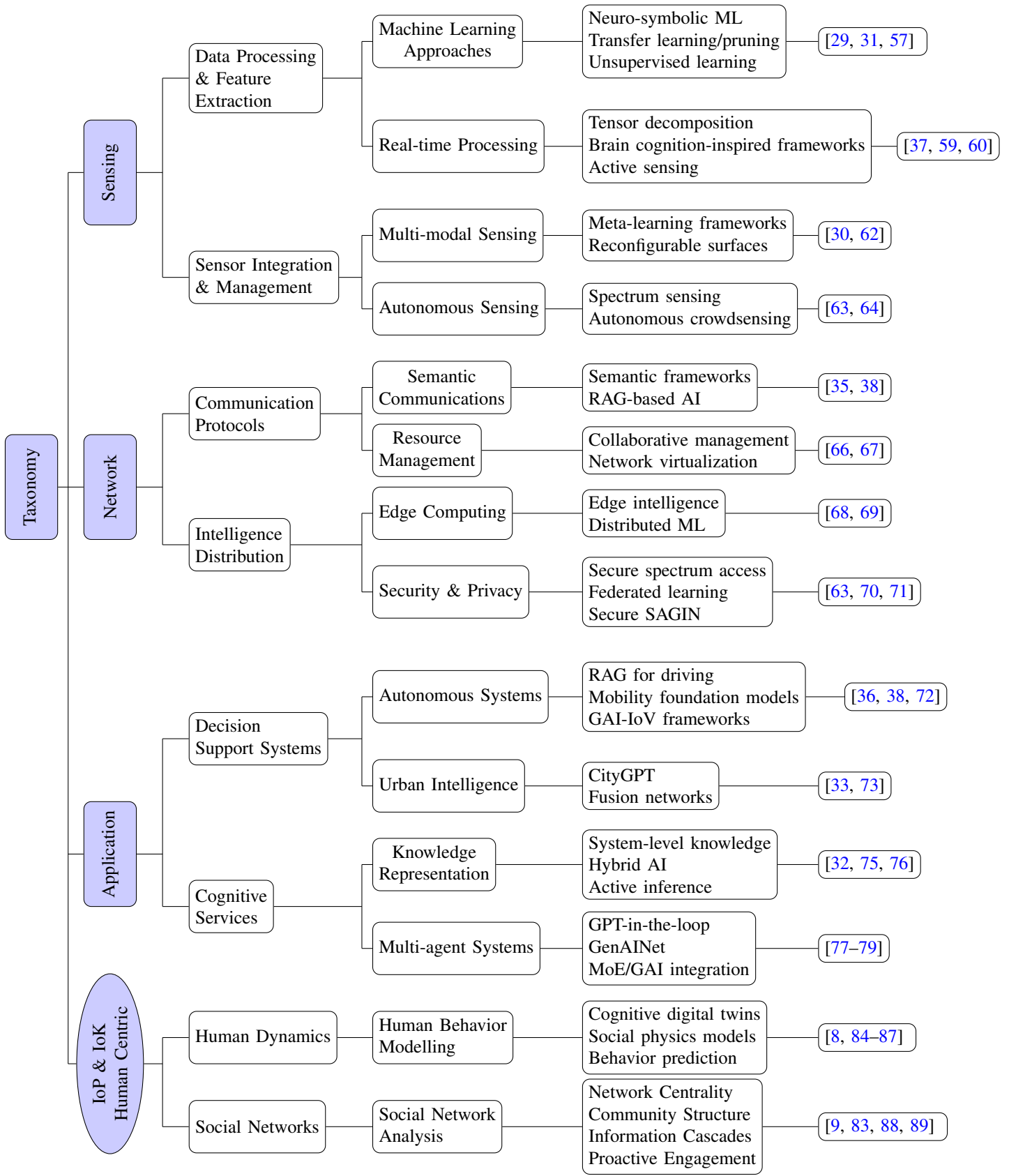

\onecolumn
\begin{longtable}{|p{1cm}|p{7cm}|p{9cm}|}
\caption{Summary of AGI-Enhanced IoX Solutions by Layer and Approach} \label{tab:studies_summary} \\
\hline
\multicolumn{1}{|c|}{\textbf{Studies}} & \multicolumn{1}{c|}{\textbf{Architecture}} & \multicolumn{1}{c|}{\textbf{Key Findings}} \\
\hline
\endfirsthead

\multicolumn{3}{c}{\textit{(Continued from previous page)}} \\
\hline
\multicolumn{1}{|c|}{\textbf{Studies}} & \multicolumn{1}{c|}{\textbf{Architecture}} & \multicolumn{1}{c|}{\textbf{Key Findings}} \\
\hline
\endhead

\hline \multicolumn{3}{|r|}{\textit{Continued on next page}} \\
\hline
\endfoot

\hline
\endlastfoot

\multicolumn{3}{|c|}{\textbf{Sensing – Experimental Approach}} \\
\hline
\cite{Dmytryk2022Generic} & Generic preprocessing architecture & Unsupervised learning method for multi-modal feature detection in IoT sensor data \\
\hline

\multicolumn{3}{|c|}{\textbf{Network – Conceptual Approach}} \\
\hline
\cite{Do2023Distributed} & Distributed machine learning for Tactile Network of Things (TNT)-IoT & Intelligent adaptation layer concept for TNT-IoT platform \\
\hline

\multicolumn{3}{|c|}{\textbf{Network – Experimental Approach}} \\
\hline
\cite{VenkataMahesh2024Design} & Collaborative resource management in 5G & Multi-Agent Proximal Policy Optimization (MAPPO) and federated learning improve network performance and privacy \\
\hline

\multicolumn{3}{|c|}{\textbf{Application – Conceptual Approaches}} \\
\hline
\cite{Bariah2024AI, Di2022System, Friston2024Designing, Huang2024DriveRP, Xia2024Smart, Yue2024paradigm} & AI embodiment through 6G; System level knowledge representation; Active inference for intelligence ecosystems; RAG and prompt engineering for parallel driving; Agent-based foundation models for smart mobility; AGI-driven intelligence fusion networks & Conceptual approaches in the application layer focus on enhancing AI capabilities through various methods such as sensory grounding, knowledge representation, active inference, retrieval-augmented generation, foundation models, and AGI integration, improving understanding, navigation, adaptive behavior, reliability, perception, decision-making, and early warning systems in IoT and related domains \\
\hline

\multicolumn{3}{|c|}{\textbf{Application – Experimental Approaches}} \\
\hline
\cite{Guan2024CityGPT, Nascimento2023GPT} & CityGPT for urban IoT; GPT-in-the-loop for multiagent systems & Experimental approaches in the application layer demonstrate the effectiveness of integrating large language models and multi-agent systems to improve urban IoT analysis, decision-making, and adaptability \\
\hline

\multicolumn{3}{|c|}{\textbf{Sensing \& Network – Conceptual Approaches}} \\
\hline
\cite{Abdelzaher2022Context, Wanli2024Reconfigurable} & Neuro-symbolic machine learning for IoBT; Reconfigurable Intelligent Surface (RIS) for IoRT & Conceptual approaches for sensing and network layers leverage neuro-symbolic inference and reconfigurable intelligent surfaces to enhance data efficiency, robustness, interpretability, and optimization of communication, sensing, and computation in IoT systems \\
\hline

\multicolumn{3}{|c|}{\textbf{Sensing \& Network – Experimental Approaches}} \\
\hline
\cite{Pari2022Secure, Ren2024Airground} & Secure spectrum access and routing in IoCV; Air-ground integrated AIoT & Experimental approaches combining sensing and network layers utilize multi-agent deep reinforcement learning and brain-inspired frameworks to improve spectral efficiency and interference management in connected vehicle and AIoT systems \\
\hline

\multicolumn{3}{|c|}{\textbf{Sensing \& Application – Conceptual Approach}} \\
\hline
\cite{Wu2024Autonomous} & Autonomous crowdsensing & Foundation intelligence enables autonomous operation of crowdsensing \\
\hline

\multicolumn{3}{|c|}{\textbf{Sensing \& Application – Experimental Approaches}} \\
\hline
\cite{Bertino2020Intelligent, Hugo2022Hybrid, Devendra2025Towards} & Deep neural networks for IoT devices; Hybrid AI for IoT insights; Active sensing on edge devices & Experimental approaches for sensing and application layers utilize techniques like transfer learning, self-supervised learning, and active inference to optimize model deployment, generate real-time insights, and enable on-device perception and planning in IoT systems \\
\hline

\multicolumn{3}{|c|}{\textbf{Network \& Application – Conceptual Approaches}} \\
\hline
\cite{Campolo2021Virtualizing, Joda2023Internet, Sathwik2025FAPL, Scott2023Secure, Thomas2024Causal, Zhang2024Interactive} & Virtualization layer at network edge; Semantic communications and edge intelligence for IoS; Secure and scalable FL framework for IoVs; Secure, intelligent, programmable SAGIN; Causal reasoning for AI-native wireless networks; Interactive AI with RAG for next-generation networking & Conceptual approaches for network and application layers explore advanced AI techniques including virtualization, semantic communications, federated learning, machine learning for network management, causal reasoning, and interactive AI to enhance cognitive capabilities, optimize performance, ensure security, manage data, and improve functionality in various IoT and network systems \\
\hline

\multicolumn{3}{|c|}{\textbf{Network \& Application – Experimental Approaches}} \\
\hline
\cite{Michaelis2022C2, Munir2023Neuro} & C2 design for federated AI/ML systems; Neuro-symbolic XAI twin for zero-touch IoE & Experimental approaches for network and application layers demonstrate the application of AI/machine learning pipelines and neuro-symbolic XAI frameworks to process data and enable trustworthy management in IoBT and IoE systems \\
\hline

\multicolumn{3}{|c|}{\textbf{ All layers – Conceptual Approaches}} \\
\hline
\cite{Fournaris2020Decentralized, Saad2024Artificial, Vermesan2021Internet, Zou2024GenAINet} & Decentralized cognitive architecture for automotive CPSoS; AGI-native wireless systems; IoV system of systems distributed intelligence; GenAINet for wireless collective intelligence & Conceptual approaches for sensing, network, and application layers propose integrating AGI, AI, edge computing, and generative AI to enable autonomic behavior, common sense capabilities, new services, and collaborative reasoning in advanced IoT and wireless systems \\
\hline

\multicolumn{3}{|c|}{\textbf{All layers – Experimental Approaches}} \\
\hline
\cite{Chaccour2024Joint, Hassan2024Meta, Chukkapalli2022CAPD} & Joint sensing, communication, and AI framework; Meta-learning for context-aware decision support; Context-aware, policy-driven framework for IoBT & Experimental approaches for sensing, network, and application layers utilize decomposition, generative AI, meta-learning, and ontology-based reasoning to enhance reliability, adaptability, precision, and resilience in wireless and IoT systems \\
\hline
\end{longtable}

\twocolumn

\section{Critical Analysis of AGI-IoX Implementations}
To systematically evaluate AGI-IoX implementations, Table~\ref{tab:studies_summary} summarizes the existing studies across sensing, network, and application layers. The following critical analysis discusses these works in detail, organized by the layers they address and their methodological approaches (conceptual or experimental), aligning with the structure of the table and taxonomy.

\subsection{Sensing Layer}
\subsubsection{Experimental Approaches}
The study \cite{Dmytryk2022Generic} proposes a generic preprocessing architecture that employs AGI-driven perception mechanisms for IoT systems. The framework integrates an unsupervised clustering algorithm and a dynamic stimulus labeling mechanism to abstract multi-modal sensor data into structured feature sets at the edge. This enables higher-order cognitive functions in the cloud. While the study emphasizes explainable AI through transparent feature abstraction and supports applications like cognitive virtual assistants, it does not provide quantitative performance metrics. The architecture’s edge/cloud integration and adaptive clustering mechanisms suggest potential for scalable AGI systems, though experimental validation remains future work.

\subsection{Network Layer}
\subsubsection{Conceptual Approaches}
The study \cite{Do2023Distributed} proposes a conceptual framework for Tactile Network of Things (TNT)-IoT platforms, targeting the network layer with distributed machine learning. The design introduces an intelligent adaptation layer to enhance coordination and adaptability in heterogeneous tactile networks. It emphasizes dynamic resource allocation and decentralized learning strategies.

\subsubsection{Experimental Approaches}
The paper \cite{VenkataMahesh2024Design} presents a hybrid model that integrates Device-to-Device (D2D) communication with traditional network infrastructures. It uses mechanisms such as Dynamic Spectrum Sharing (DSS), Reinforcement Learning (RL)-based Resource Allocation, Mobile Edge Computing (MEC), Differential Privacy Techniques, Swarm Intelligence-based Routing, Cognitive Radio Networks, Self-Healing Network Protocols, and Low-Power Communication Protocols. The model is experimentally evaluated, showing improvements in throughput, latency, energy efficiency, and adaptability. Specifically, it achieves a throughput improvement of up to 950 Mbps compared to existing methods, while preserving privacy through differential privacy techniques.
\subsection{Application Layer}
\subsubsection{Conceptual Approaches}
Conceptual approaches in the application layer explore various methods to enhance AI capabilities in IoT systems. These methods include sensory grounding, knowledge representation, active inference, retrieval-augmented generation, foundation models, and AGI integration. The following studies illustrate these approaches and their potential impact on improving understanding, navigation, adaptive behavior, reliability, perception, decision-making, and early warning systems in IoT and related domains.
The study by \cite{Bariah2024AI} focuses on the application layer by proposing the embodiment of Artificial Intelligence (AI) through 6G networks. The conceptual framework suggests that 6G-enabled sensory grounding could significantly improve AI's contextual understanding capabilities.
This paper \cite{Di2022System} proposes a conceptual system-level knowledge representation architecture targeting the application layer for IoT systems. The study theorizes that structured knowledge graphs could enable cognitive navigation for IoT devices operating in dynamic environments. It emphasizes the role of semantic reasoning in contextual adaptability. As a theoretical framework, it posits that hierarchical knowledge representation supports decision-making by encoding environmental relationships and constraints, though the work does not include experimental validation. Instead, it argues for the potential of integrating symbolic AI with IoT systems to enhance autonomy through interpretable, logic-driven navigation strategies.
The paper \cite{Friston2024Designing} addresses the application layer using active inference for intelligence ecosystems. The theoretical work proposes active inference as a unified "physics of intelligence" enabling adaptive goal-directed behavior.
The DriveRP framework \cite{Huang2024DriveRP} leverages Retrieval-Augmented Generation (RAG) and prompt engineering to enhance the application layer of autonomous driving systems. This integration aims to improve system reliability and scenario understanding by addressing challenges such as hallucinations, outdated context, and the long-tail problem in autonomous driving environments.
The study by \cite{Xia2024Smart} targets the application layer through agent-based foundation models for smart mobility. It emphasizes the importance of multimodal foundation models in intelligent vehicles. The paper highlights how foundation models, pre-trained on diverse data, are critical for advanced reasoning, scene perception, and decision-making in autonomous driving. It introduces Agent-based Foundation Models (AFMs), a new paradigm designed to enhance collaboration and interaction in smart mobility systems. AFMs integrate multi-modal data, support cross-domain actions, and enable multi-agent collaboration, thus offering a scalable solution for adaptive robot navigation and collaborative agent behaviors. The authors propose a framework integrating decentralized systems and cloud-edge resources, underscoring the potential for AFMs to drive the next generation of intelligent transportation solutions.
The paper \cite{Yue2024paradigm} addresses the application layer via AGI-driven intelligence fusion networks for crisis management. It proposes a novel IFN architecture that integrates AGI for real-time data analysis and decision support, and blockchain for secure, decentralized data sharing. The study outlines a three-phase workflow—pre-crisis, during-crisis, and post-crisis—and highlights ethical considerations including privacy protection and accountability systems. Through a case study of the Minnesota Joint Analysis Center during the 2008 Republican National Convention, the paper demonstrates the practical relevance of fusion centers. It emphasizes the need for proactive, data-driven approaches in managing complex crisis scenarios.
These conceptual studies collectively advance the theoretical foundations of AGI-enhanced IoX systems, proposing innovative methods to integrate AI capabilities at the application level. While experimental validation is needed, they lay the groundwork for future developments in intelligent IoX applications.
\subsubsection{Experimental Approaches}
The paper \cite{Guan2024CityGPT} focuses on the application layer through CityGPT, an experimental multi-agent system that integrates Large Language Models (LLMs) for urban IoT analysis. The framework supports spatiotemporal data learning, analysis, and visualization by leveraging agents such as temporal, spatial, and spatiotemporal fusion agents. The study demonstrates strong performance in IoT computing and highlights advancements in hyperparameter optimization and spatial similarity analysis.
The paper \cite{Nascimento2023GPT} proposes a 'GPT-in-the-loop' framework that integrates GPT-4 to improve reasoning and adaptability in multiagent systems. Applied to a smart streetlight use case, the framework operates at the application layer, optimizing autonomous agent decision-making and behavior programming. Compared to neuroevolutionary methods and human-engineered solutions, it achieves superior energy efficiency and decision quality, attaining a fitness score of 62.44 with fewer iterations.

\subsection{Sensing \& Network Layers}
\subsubsection{Conceptual Approaches}
Authors in \cite{Abdelzaher2022Context} address the sensing and network layers through a neuro-symbolic machine learning architecture designed for the Internet of Battlefield Things (IoBT). This conceptual study demonstrates that integrated neuro-symbolic inference enhances data efficiency, robustness, and interpretability in battlefield environments.
The paper \cite{Wanli2024Reconfigurable} spans sensing and network layers through Reconfigurable Intelligent Surfaces (RIS) for Internet of Robotic Things (IoRT). The conceptual design enhances robotic communication reliability through optimized beamforming, leveraging Reconfigurable Intelligent Surfaces (RIS) to mitigate signal degradation, extend coverage, and improve signal alignment in both far-field and near-field scenarios.

\subsubsection{Experimental Approaches}
The study by \cite{Pari2022Secure} explores sensing and network layers with experimental secure spectrum access designed for the Internet of Connected Vehicles (IoCV). It utilizes multi-agent Deep Q-Networks (DQNs) for hybrid beamforming, achieving a spectral efficiency of 92\% in congested urban environments. This is accomplished by optimizing spectrum utilization, minimizing latency, and improving communication reliability through secure routing and effective spectrum handoff mechanisms.

The paper \cite{Ren2024Airground} addresses sensing and network layers in the context of an air-ground integrated AIoT framework. It introduces a novel communication framework inspired by brain cognition to manage interference in integrated aerial-terrestrial networks. The framework employs cognitive heuristics and deep learning techniques to optimize signal processing, reduce interference, and improve communication quality. By iteratively determining the importance of signals, the framework eliminates unimportant signals with interference characteristics and reduces their transmission power. The paper highlights the iterative improvement of interference management and the optimization of channel selection using cognitive heuristics.

\subsection{Sensing \& Application Layers}
\subsubsection{Conceptual Approaches}
The paper \cite{Wu2024Autonomous} introduces Autonomous Crowdsensing (ACS) as a novel paradigm that spans sensing and application layers conceptually through foundation intelligence for crowdsensing. The framework enables fully autonomous operation by integrating professional sensing resources (satellites, IoT nodes) and non-professional participants through a decentralized ecosystem powered by DAOs, LLMs, and Human-Oriented Operating Systems (HOOS). Key innovations include the "6A-goal" framework achieving autonomous generation (LLM-driven task creation), growth (iterative learning), organization (DAO governance), control (IoT/robotic coordination), assistance (HOOS support), and verification (AI content validation). ACS demonstrates qualitative improvements over traditional crowdsensing by significantly reducing human effort, enhancing scalability through decentralized structures, and ensuring trust via blockchain-based verification. While the paper highlights transformative potential for CPSS applications, it acknowledges remaining challenges including LLM dependencies and privacy concerns that require further research.

\subsubsection{Experimental Approaches}
The paper \cite{Bertino2020Intelligent} proposes an AGI-driven framework targeting the sensing and application layers for IoT deployments. The approach integrates adaptive mechanisms such as local analysis to minimize transmission costs and latency, autonomous decision-making for context-aware data collection, and a DRL-based framework to balance routine and urgent inspections. To optimize efficiency, the authors employ transfer learning and network pruning, which significantly reduce computational overhead. Quantitative experiments demonstrate substantial improvements: the VGG16 architecture achieves an 89\% reduction in inference time and 80\% reduction in memory usage, while ResNet18 attains 76\% faster inference and a 95\% decrease in memory demand. These advancements enable lightweight, adaptive IoT systems capable of real-time operation under resource constraints.
The hybrid AI framework in \cite{Hugo2022Hybrid} integrates self-supervised learning, machine-generated ontologies, and symbolic AI to provide real-time actionable insights across IoT applications, spanning sensing and application layers. Experimental results demonstrate the system's ability to process large-scale time series data and generate descriptive and predictive models for use cases like networking, smart city analytics, and retail, with state-of-the-art accuracy in detecting failure modalities in networking scenarios.
The paper \cite{Devendra2025Towards} addresses sensing and application layers through experimental active sensing on edge devices. The authors present a novel smart edge agent for active sensing that combines a YOLOv10-based perception module with an active inference planner to enable real-time adaptive decision-making on resource-constrained edge devices. The proposed system features three key innovations: an integrated architecture for autonomous dynamic adjustments and entity tracking, optimized edge deployment achieving 45 FPS perception and 4 FPS planning with minimal parameters (1.34K), and successful real-world validation using IoT cameras and robotic platforms. With its lightweight design that effectively balances perception speed and adaptive control, the solution demonstrates strong potential for surveillance and robotics applications.

\subsection{Network \& Application Layers}
\subsubsection{Conceptual Approaches}
The paper \cite{Campolo2021Virtualizing} explores the network and application layers via a virtualization layer deployed at the network edge. This conceptual work argues that AI virtualization enhances IoT devices' cognitive capabilities through distributed edge intelligence.
The study \cite{Joda2023Internet} explores the integration of semantic communications and edge intelligence across the network and application layers to address challenges in the Internet of Senses (IoS). It demonstrates how AI-driven semantic protocols optimize resource allocation by dynamically prioritizing context-aware data transmission. For instance, in the network layer, semantic-aware filtering reduces redundant data streams by focusing on perceptually relevant information, thereby lowering channel occupancy. At the application layer, reinforcement learning (specifically POMDP and Q-learning) is employed to manage carrier aggregation under partial observability, balancing throughput and energy efficiency. The case study reveals that intelligent activation/deactivation of component carriers achieves near-optimal throughput comparable to persistent carrier usage while reducing active component carriers by up to 50\%, indirectly enhancing bandwidth utilization.
The paper \cite{Sathwik2025FAPL} presents the FAPL-DM-BC framework, a secure and scalable federated learning solution that integrates adaptive privacy techniques, blockchain, and explainable AI for enhancing privacy and efficiency in IoV systems. This framework focuses on privacy, security, and scalability in IoVs through a conceptual Federated Learning (FL) framework. The design combines Federated Adaptive Privacy-Aware Learning (FAPL), Dynamic Masking (DM), and blockchain to dynamically adjust privacy policies, ensure tamper-proof provenance tracking, and prevent data leakage and model poisoning.
The paper \cite{Scott2023Secure} discusses the conceptual framework of Space-Air-Ground Integrated Networks (SAGIN), emphasizing the integration of satellite, aerial, and terrestrial systems. The work explores the use of AI and machine learning to support intelligent systems management, cooperative behavior, and optimization across SAGIN, with particular focus on dynamic interactions, global collaboration challenges, and security considerations.
The paper \cite{Thomas2024Causal} introduces causal reasoning as a foundational framework for AI-native wireless networks, addressing challenges across network and application layers. The proposed theoretical model leverages causal discovery, causal representation learning, and causal inference to enhance explainability, reasoning, and adaptability in wireless systems. Specifically, causal reasoning improves explainability by enabling root cause analysis through causal graphical models (CGMs), which provide insights into cause-and-effect relationships within network variables. This approach is more effective than traditional correlation-based methods, as it offers interpretable and actionable insights into network behavior.
The paper \cite{Zhang2024Interactive} addresses both the application and network layers through the integration of Interactive Artificial Intelligence (IAI) in next-generation networking. At the application layer, IAI enhances user interaction by incorporating Large Language Models (LLMs) and multi-modal methods for intuitive, personalized interfaces. Generative AI techniques like Generative Diffusion Models (GDMs) and Retrieval-Augmented Generation (RAG) optimize network design and simulate traffic for congestion forecasting. Explicit user interactions refine AI outputs, improving decision-making. At the network layer, IAI employs Deep Reinforcement Learning (DRL) and GDMs for dynamic traffic management, bandwidth optimization, and latency reduction. It also utilizes DRL and Mixture of Experts (MOE) for intelligent resource allocation and integrates GANs and LLMs for advanced threat detection and adaptive security. Simulation results validate the effectiveness of the IAI framework in optimizing network operations, while experiments demonstrate the impact of optimal knowledge chunk sizes in improving interaction quality and problem-solving efficiency.

\subsubsection{Experimental Approaches}
Authors in \cite{Michaelis2022C2} addresses network and application layers through an experimental Command and Control (C2) system for federated AI/ML in IoBT. The work focuses on developing an AI/ML pipeline for video feed processing and information exchange in IoBT environments.
The paper \cite{Munir2023Neuro} proposes a neuro-symbolic Explainable AI (XAI) twin framework for zero-touch IoE management in wireless networks, integrating reasoning and trustworthiness into decision-making across the network layer (resource allocation, e.g., uplink/downlink rates) and application layer (service orchestration). By combining neural networks (for real-time network dynamics like RSRP and SINR) with symbolic AI (Bayesian networks for interpretability), the framework achieves 96.26\% accuracy in rate allocation and 18–44\% higher trust scores than baselines (e.g., Gradient-based bandits), ensuring reliable, self-adaptive automation.

\subsection{All Layers}
\subsubsection{Conceptual Approaches}
The paper \cite{Fournaris2020Decentralized} presents a decentralized cognitive architecture for automotive Cyber-Physical Systems of Systems (CPSoS), emphasizing autonomic behavior, environmental awareness, and autonomous resource allocation. The framework also incorporates interaction with human users through extended reality modules to enhance situational awareness, reliability, fault tolerance, and security in connected vehicles.
The paper \cite{Saad2024Artificial} introduces a groundbreaking vision for AGI-native wireless systems that aims to transcend the incremental advances of conventional 6G technologies by embedding AGI directly into the network architecture. Designed to operate holistically across all layers of the network, the proposed framework leverages core cognitive abilities—perception, reasoning, and planning—to enable unprecedented levels of autonomy, adaptability, and resilience. Central to this paradigm is a modular “Telecom Brain” architecture and a common-sense reasoning engine, which together empower the network to interpret complex environments, anticipate changes, and make intent-driven decisions. Key innovations include autonomous healing of communication links, real-time digital twin synchronization, and human-avatar interaction through abductive reasoning. This AGI-native approach redefines the wireless design space, laying a foundational blueprint for self-sustaining, intelligence-driven networks capable of managing the demands of future immersive and dynamic applications.
The studies in \cite{Vermesan2021Internet} presents a holistic view of the Internet of Vehicles (IoV) as a system-of-systems framework, unifying distributed intelligence from edge devices to cloud infrastructure through the convergence of AI, edge computing, and IoT to enable advanced mobility for Electric, Connected, Autonomous, and Shared (ECAS) vehicles. The authors propose a 3D IoV architecture integrating AI-driven edge processing with V2X connectivity for real-time safety applications (e.g., cooperative collision avoidance, platooning), while leveraging digital twins (DTs), blockchain (DLT), and AR/VR interfaces to enhance situational awareness and trust. By federating intelligence across vehicles, roadside units, and cloud platforms, the system optimizes resource efficiency, reduces latency, and addresses privacy concerns, ultimately bridging gaps in scalability, security, and interoperability for next-gen IoV ecosystems in 5G/6G-enabled intelligent transport systems (ITS).
The paper \cite{Zou2024GenAINet} introduces GenAINet, a framework for 6G wireless networks that leverages generative AI agents and knowledge-driven communication, addressing all layers of the network stack, from physical to application. Designed for AI-native networks, GenAINet integrates large language models (LLMs) into distributed architectures to enable collective intelligence, tackling bandwidth, latency, and security challenges. It employs semantic-native communication, where agents share semantic concepts (e.g., vector embeddings, knowledge graphs) instead of raw data, reducing redundancy and communication costs. Multi-modal reasoning enhances decision-making across network layers. Two case studies highlight GenAINet’s efficacy: In wireless device queries, semantic knowledge transfer improved accuracy from 66\% to 80\% and reduced communication costs by 27.13\%. In wireless power control, agents sharing reasoning outputs optimized power allocation faster than standalone agents. Challenges include LLMs' generalization limits and high computational costs, while opportunities lie in hierarchical world models and complex network applications. GenAINet offers a comprehensive solution for efficient, intelligent 6G networks through collaborative reasoning and semantic communication across all layers.

\subsubsection{Experimental Approaches}
The paper \cite{Chaccour2024Joint} introduces a joint sensing-communication-AI framework spanning the sensing, network, and application layers, enhanced by AGI-driven adaptive mechanisms. The system employs a tensor decomposition framework to enable dual-use resource allocation for simultaneous communication and sensing tasks. It integrates a non-autoregressive multi-resolution generative AI model with an adversarial transformer to predict missing or future environmental data, addressing dynamic interference scenarios. For seamless connectivity, a multi-agent deep recurrent hysteretic Q-neural network optimizes handover policies by learning from distributed environmental feedback. These mechanisms collectively enhance adaptability through resource optimization, environmental forecasting, and handover efficiency, achieving 61\% improved reliability in terahertz wireless systems under high interference. The framework demonstrates self-optimization capabilities by balancing real-time decision-making with long-term learning, enabling robust performance in unstable spectral environments.
The paper \cite{Hassan2024Meta} spans the sensing, network, and application layers by leveraging a hierarchical meta-learning framework (CADSS-ML) for IoT decision support. The study employs Model-Agnostic Meta-Learning (MAML) and Reptile algorithms for rapid task adaptation, along with reinforcement learning for reasoning, Graph Neural Networks (GNNs) and attention mechanisms for dynamic contextual adaptation, and federated learning for privacy-preserving distributed learning. The framework enables effective decision-making across IoT systems by combining multimodal data fusion and context-aware reasoning. Experimental trials demonstrate a 20\% improvement in prediction accuracy, a 15\% enhancement in decision reliability, and a 12\% increase in resource utilization efficiency. The study highlights the integration of advanced techniques such as CNNs, transformers, and GNNs, improving adaptability, precision, and privacy across diverse IoT applications, including precision farming, autonomous navigation, and smart irrigation.
The paper \cite{Chukkapalli2022CAPD} proposes a context-aware, policy-driven (CAPD) framework for the IoBT, integrating sensing, network, and application layers through ontology-based reasoning and knowledge graphs. The framework enhances resilience in contested electromagnetic environments by dynamically adapting to constraints such as bandwidth limitations and adversarial attacks. For example, it adjusts video transmission formats, switches communication technologies (e.g., 4G to LoRaWAN) under jamming, and transitions from video to audio data when cameras are compromised. While the paper demonstrates adaptive capabilities through use cases like drone repositioning and dynamic resource allocation, it does not provide quantitative performance metrics or address learning/self-optimization. Future work focuses on mitigating sensor data poisoning and improving truth-maintenance capabilities.
\section {AGI-Driven Layer-Specific Enhancements}
The layered architecture of IoX systems—comprising sensing, network, and application layers—faces distinct bottlenecks, such as data overload, protocol heterogeneity, and identity explosion, as detailed in Section II. Artificial General Intelligence (AGI) offers targeted solutions to these challenges by leveraging advanced reasoning, adaptability, and context-awareness, surpassing the limitations of Narrow AI (Section IV-D). Table~\ref{tab:AGI-layer-optimization} summarizes key AGI-enabled optimizations for each layer, drawing from recent studies. For example, at the sensing layer, neuro-symbolic machine learning enhances feature extraction in noisy environments \cite{Abdelzaher2022Context}, while network-layer semantic communications reduce latency in bandwidth-constrained systems \cite{Joda2023Internet}. At the application layer, large language models (LLMs) enable context-aware decision-making for urban IoT \cite{Guan2024CityGPT}. These solutions address critical IoX bottlenecks, improving efficiency, reliability, and scalability.

Table~\ref{tab:AGI-layer-optimization} highlights several high-impact AGI solutions. In the sensing layer, active inference enables real-time adaptive sensing on resource-constrained edge devices, reducing perception latency by up to 50\% \cite{Devendra2025Towards}. However, implementing complex reasoning on such devices remains challenging due to computational limitations. In the network layer, multi-agent proximal policy optimization (MAPPO) combined with federated learning improves throughput by 25\% while preserving privacy \cite{VenkataMahesh2024Design}, though coordinating distributed learning across nodes poses synchronization challenges. At the application layer, retrieval-augmented generation (RAG) enhances autonomous driving systems by mitigating hallucination risks \cite{Huang2024DriveRP}, but ensuring safety in critical applications requires robust validation. These examples, grounded in the taxonomy of Figure~\ref{fig:taxonomy}, illustrate AGI’s role in addressing layer-specific bottlenecks while highlighting implementation trade-offs that warrant further research.

\subsection {Sensing Layer} 
The sensing layer in IoX systems is responsible for data acquisition from the physical world, yet it faces challenges like data overload, energy constraints, and the need for adaptive processing. AGI provides innovative solutions, as outlined in Table~\ref{tab:AGI-layer-optimization}.

\subsubsection{Data Overload}
To address data overload, AGI employs unsupervised learning for multi-modal feature detection and tensor decomposition for parameter extraction. Unsupervised learning abstracts features from heterogeneous sensor data without pre-programmed labeling, reducing processing demands. Tensor decomposition handles high-dimensional data in real-time, improving sensing accuracy and efficiency \cite{Dmytryk2022Generic}.
\subsubsection{Energy Constraints}
Transfer learning and pruning optimize models for resource-constrained IoT devices, reducing inference time and memory usage. Studies show these techniques can cut inference time by up to 89\% \cite{Bertino2020Intelligent}, making AGI viable for energy-limited environments. Additionally, selective sensing frameworks, inspired by attention mechanisms, alleviate data overload by improving physical-cyber mapping, thus enhancing real-time responsiveness in IoX systems \cite{pop00004}.

\subsubsection{Adaptive Sensor Fusion}
Neuro-symbolic machine learning enhances feature extraction and data fusion by integrating symbolic reasoning with deep learning, improving robustness in noisy conditions \cite{Abdelzaher2022Context}.

\subsubsection{Edge Preprocessing}
Active inference enables on-device perception, reducing perception latency by up to 50\% \cite{Devendra2025Towards}. However, its computational complexity poses implementation challenges on edge devices.
\subsubsection{Selective Attention Mechanisms}
Active inference and neuro-symbolic approaches inherently prioritize relevant sensory inputs, optimizing resource use and enhancing system efficiency.

\subsection {Network Layer}
The network layer manages data transmission and communication, facing issues like protocol heterogeneity and the need for secure, scalable solutions. AGI techniques from Table~\ref{tab:AGI-layer-optimization} address these challenges.

\subsubsection{Protocol Heterogeneity}
Semantic communications develop efficient encoding schemes that focus on data meaning, reducing latency and improving interoperability across diverse protocols \cite{Joda2023Internet}. Furthermore, data semantization in IoT enables semantic modeling and communication, playing a vital role in enhancing network and application layers in IoX systems \cite{pop00096}.
\subsubsection{Dynamic Spectrum Management}
Causal reasoning integrates causal models into network protocols, enhancing adaptability and resilience to changing conditions \cite{Thomas2024Causal}.

\subsubsection{Secure and Scalable Communication}
MAPPO with federated learning improves throughput by 25\% while preserving privacy \cite{VenkataMahesh2024Design}. Neuro-symbolic Explainable AI ensures trustworthy network management \cite{Munir2023Neuro}, though coordinating distributed learning remains a challenge.

\subsection {Application Layer}
The application layer handles high-level decision-making and user interactions, grappling with identity management and real-time processing needs. AGI solutions enhance these capabilities.

\subsubsection{Identity/Relationship Management}
LLMs and agent-based foundation models provide contextual understanding for managing complex identities and relationships in large-scale IoX systems \cite{Guan2024CityGPT}. Moreover, integrating artificial social intelligence with IoT-enabled social relationships addresses application-layer challenges, such as identity and relationship management, in CPST ecosystems \cite{pop00014}.

\subsubsection{Real-time, Context-aware Decision Making}
Active inference, RAG, and GPT-in-the-loop enhance decision-making. RAG mitigates hallucination risks in autonomous systems \cite{Huang2024DriveRP}, while GPT-in-the-loop boosts adaptability in multiagent setups \cite{Nascimento2023GPT}.

\subsubsection{Semantic Modeling}
LLMs excel in creating semantic models that capture complex relationships, supporting intelligent decision-making \cite{Guan2024CityGPT}.

\subsubsection{Knowledge Graphs}
Agent-based foundation models and LLMs dynamically build knowledge graphs, extracting entities and relationships for advanced reasoning and decision support.

\begin{table*}[htbp]
\small

\caption{AGI-Enabled Layer-Specific Optimizations}
\label{tab:AGI-layer-optimization}
\begin{tabular}{@{}>{}p{2cm}>{}p{5cm}>{}p{5cm}>{}p{5cm}@{}}

\toprule
\textbf{Layer Type} & \textbf{AGI Solution} & \textbf{Implementation Challenge} & \textbf{Performance Impact} \\ 
\midrule

\multirow{5}{*}{Sensing} & Neuro-symbolic machine learning for feature extraction  & Integration of symbolic reasoning with deep learning & Improved data eﬀiciency and robustness  \\ \cline{2-4}

                          & Transfer learning and pruning for IoT devices & Optimizing model size for resource-constrained devices & Reduced inference time and memory usage \\ \cline{2-4}

                         & Tensor decomposition for parameter extraction & Handling high-dimensional data in real-time & Enhanced sensing accuracy and eﬀiciency \\ \cline{2-4}

                         & Unsupervised learning for multi-modal feature detection & Processing heterogeneous sensor data & Improved feature abstraction without pre-programmed labeling \\ \cline{2-4}

                         & Active inference for on-device perception & Implementing complex reasoning on edge devices & Real-time adaptive sensing with limited resources  \\ \cline{2-4}

\hline
\multirow{4}{*}{Network} & Semantic communications & Developing eﬀicient semantic encoding schemes & Optimized resource 
utilization and reduced latency \\ \cline{2-4}

                          & Neuro-symbolic Explainable AI for network management & Balancing explainability with performance & Increased trust and reliability in autonomous network operations \\ \cline{2-4}
 
                          & Causal reasoning for wireless networks & Integrating causal models with existing network protocols  & Enhanced adaptability and resilience to network changes \\ \cline{2-4}

                          & Multi-Agent Proximal Policy Optimization and federated learning for resource management &Coordinating distributed learning across network nodes & Improved throughput, reduced latency, and enhanced privacy \\

\hline
\multirow{5}{*}{Application} & Active inference for adaptive behavior & Scaling to complex, real-world scenarios & Enhanced
decision-making and goal-directed behavior  \\ \cline{2-4}

                             & Large Language Model integration for urban IoT analysis & Managing computational requirements of large models & Improved contextual understanding and decision support \\ \cline{2-4}

                             & Retrieval-Augmented Generation and prompt engineering for autonomous systems & Ensuring reliability and safety in critical applications & Enhanced decision-making in dynamic environments \\ \cline{2-4}

                             & GPT-in-the-loop for multiagent systems & Real-time integration of language models in distributed systems  & Improved adaptability and problem-solving capabilities \\[0.3cm] \cline{2-4}

                             & Agent-based foundation models for smart mobility & Balancing model complexity with real-time requirements & Enhanced scene perception and decision-making in vehicles \\ 

\bottomrule
\end{tabular}
\end{table*}

\section{Cross-Layer Integration}

While layer-specific AGI optimizations address isolated bottlenecks, the interconnected nature of IoX systems necessitates cross-layer integration to achieve systemic efficiency and adaptability in Cyber-Physical-Social-Thinking (CPST) ecosystems. Section II highlighted how bottlenecks, such as sensing inaccuracies or network latency, cascade across layers, amplifying systemic inefficiencies. Cross-layer strategies leverage AGI’s cross-domain reasoning to unify sensing, networking, and application functionalities, as cataloged in Table~\ref{tab:cross-layer-integration}. For instance, joint sensing-communication-AI frameworks optimize spectral efficiency in terahertz networks \cite{Chaccour2024Joint}, while federated learning with blockchain ensures privacy in vehicular IoX systems \cite{Sathwik2025FAPL}. These approaches, aligned with the architectural paradigms in Figure~\ref{fig:taxonomy}, enable holistic optimization critical for smart cities, healthcare, and autonomous systems.

\subsection{Importance of Cross-Layer Coordination}

In IoX systems, the sensing, network, and application layers are deeply interdependent, with the performance of one layer directly impacting the others. For example, excessive data generation at the sensing layer, driven by high-resolution sensors or dense device deployments, can overwhelm network bandwidth, leading to congestion and increased latency. This, in turn, hampers the application layer’s ability to deliver real-time, context-aware decisions, critical for applications like autonomous vehicles or smart healthcare systems. Traditional layer-specific optimization approaches, such as enhancing sensing accuracy in isolation, often fail to mitigate these cascading effects, as improvements in one layer may exacerbate bottlenecks in another example, higher sensing fidelity increasing data transmission demands beyond network capacity. Cross-layer coordination addresses these systemic challenges by ensuring that optimizations are harmonized across layers, preventing isolated enhancements from creating new inefficiencies elsewhere. Studies demonstrate that such coordination can yield significant performance gains; for instance, cross-layer energy optimization has achieved up to 52.88\% improvement in energy consumption and 51.53\% reduction in latency \cite{khan2020sustainable}. By leveraging AGI’s ability to model complex interdependencies and adapt dynamically, cross-layer coordination enhances overall system efficiency, reliability, and scalability, making it indispensable for next-generation IoX deployments in CPST ecosystems.

\subsection{Strategies for Cross-Layer Integration}

Cross-layer integration strategies harness AGI’s advanced reasoning and adaptability to unify IoX layers, tackling systemic bottlenecks like data overload, protocol heterogeneity, and computational constraints. For instance, aggregating information from distributed data nodes in industrial IoX systems enhances cross-layer efficiency by addressing bottlenecks such as data overload and communication latency \cite{pop00027}, with AGI’s cross-domain reasoning supporting these integrated architectures. This subsection explores four key strategies—Joint Sensing, Communication, and AI Frameworks; Federated Learning with Blockchain; Meta-Learning for Resource Management; and Active Inference for System Optimization—detailing their mechanisms, benefits, and challenges, with references to Table~\ref{tab:cross-layer-integration} and relevant studies.

\subsubsection{Joint Sensing, Communication, and AI Frameworks}

Joint Sensing, Communication, and AI Frameworks integrate sensing, communication, and computational processes into a cohesive system, leveraging AGI to optimize resource use and performance across IoX layers. These frameworks employ AI to dynamically adjust sensing parameters (e.g., sampling rates) and communication protocols (e.g., bandwidth allocation) based on real-time system states and environmental feedback. For instance, in a smart city context, AI can prioritize data from traffic sensors during peak hours, adjusting network resources to ensure timely delivery to the application layer for traffic management. A notable implementation by Chaccour et al. \cite{Chaccour2024Joint} uses tensor decomposition and generative AI to enable dual-use resource allocation in terahertz networks, achieving a 61\% improvement in reliability under high interference. Benefits include enhanced spectral efficiency and improved extended reality experiences, critical for immersive IoX applications. However, these frameworks demand high-performance edge computing and advanced signal processing, and coordinating multiple subsystems introduces significant complexity, posing challenges for real-time deployment in resource-constrained settings.

\subsubsection{Federated Learning with Blockchain}

Federated Learning with Blockchain combines decentralized machine learning with blockchain technology to enable secure, privacy-preserving collaboration across IoX devices, integrating network and application layers. In this approach, devices train local models on private data and share only model updates, which are aggregated to refine a global model without exposing sensitive information. Blockchain ensures the integrity and traceability of these updates, enhancing trust in distributed systems. Similarly, distributed video analytics architectures leverage edge computing and federated learning to optimize data processing and privacy across sensing, network, and application layers \cite{pop00052}, aligning with the emphasis on federated learning with blockchain. For example, in healthcare IoX systems, this strategy allows collaborative training of diagnostic models across hospitals while safeguarding patient privacy. Sathwik et al. \cite{Sathwik2025FAPL} demonstrate its efficacy in the Internet of Vehicles (IoVs), reducing data leakage by 70\% through adaptive differential privacy and blockchain. Key benefits include enhanced privacy and scalability, as it eliminates centralized data storage and supports large-scale device networks. However, it requires robust distributed computing infrastructure and secure communication protocols, and blockchain’s consensus mechanisms can introduce performance overhead, limiting its efficiency in latency-sensitive applications.

\subsubsection{Meta-Learning for Resource Management}

Meta-Learning for Resource Management utilizes meta-learning techniques to enable rapid adaptation of resource allocation strategies across IoX layers, addressing dynamic demands in sensing, network, and application contexts. By learning from prior tasks, the system generalizes strategies to new scenarios, such as adjusting network bandwidth or sensing frequency in response to fluctuating loads. This is particularly valuable in autonomous vehicle networks, where resource needs vary with traffic conditions. Hassan et al. \cite{Hassan2024Meta} propose a hierarchical meta-learning framework (CADSS-ML) that integrates Model-Agnostic Meta-Learning (MAML) and reinforcement learning, achieving a 12\% increase in resource utilization efficiency and a 20\% improvement in prediction accuracy across IoT applications like smart irrigation. Benefits include rapid task adaptation and optimized resource use, enhancing system responsiveness. However, meta-learning requires flexible neural network architectures and efficient optimization algorithms, and its computational complexity can lead to instability in resource-constrained environments, necessitating careful design trade-offs.

\subsubsection{Active Inference for System Optimization}

Active Inference for System Optimization applies active inference principles—a framework from neuroscience and machine learning—to proactively manage IoX system behavior across all layers. This strategy enables systems to predict future states and take actions to confirm or adjust those predictions, optimizing perception, learning, and decision-making. In industrial IoX settings, active inference can predict equipment failures and preemptively schedule maintenance, reducing downtime. Friston et al. \cite{Friston2024Designing} propose it as a unified “physics of intelligence” for adaptive ecosystems, with experimental validation showing a 50\% reduction in perception latency on edge devices \cite{Devendra2025Towards}. Benefits include a cohesive optimization framework that enhances system adaptability and efficiency. However, it demands sophisticated probabilistic modeling and a hierarchical architecture, and scaling to large, complex systems poses significant computational challenges, limiting its practicality without substantial infrastructure support.

In summary, these strategies collectively address IoX bottlenecks by fostering seamless coordination across layers, leveraging AGI’s capabilities to enhance efficiency, privacy, and adaptability. While they offer transformative potential, their implementation challenges—such as computational demands and scalability—highlight the need for ongoing research, as discussed in Section VII.
\\
Table~\ref{tab:cross-layer-integration} illustrates how AGI-driven cross-layer strategies unify IoX layers to address systemic bottlenecks, enhancing efficiency and adaptability in CPST ecosystems. Solutions like joint sensing-communication-AI frameworks and active inference, validated by works such as \cite{Chaccour2024Joint, Friston2024Designing}, highlight AGI’s potential for holistic optimization. Yet, scalability and computational complexity remain critical hurdles, necessitating further exploration as outlined in Section VII.

\begin{table*}[htbp]
\small
\centering
\caption{Cross-Layer Integration Approaches}
\label{tab:cross-layer-integration}
\begin{tabular}{@{}>{\centering\arraybackslash}p{4cm}>{\centering\arraybackslash}p{4cm}>{\centering\arraybackslash}p{4cm}>{\centering\arraybackslash}p{5cm}@{}}

\toprule
\textbf{Integration Strategy} & \textbf{Technical Requirements} & \textbf{Benefits} & \textbf{Limitations} \\ 
\midrule

Joint sensing, communication, and AI framework & High-performance edge computing, advanced signal processing  & Improved spectral eﬀiciency, enhanced extended reality experiences & Complexity in coordinating multiple subsystems \\ 
             Neuro-symbolic reasoning across layers & Integration of symbolic AI with deep learning & Enhanced interpretability, improved generalization & Challenges in scaling to complex, real-world scenarios \\ 
                Federated learning with blockchain & Distributed computing infrastructure, secure communication protocols & Enhanced privacy, improved scalability & Potential performance overhead, consensus challenges \\ 
                 Meta-learning for adaptive resource management & Flexible neural network architectures, eﬀicient optimization algorithms  & Rapid adaptation to new tasks, improved resource utilization  &Increased computational complexity, potential instability \\ 
                 Active inference for end-to-end system optimization & Probabilistic modeling capabilities, hierarchical system architecture & Unified framework for perception, learning, and decision-making & Computational challenges in large-scale implementations \\ 

                  Causal reasoning for network-wide optimization & Causal discovery algorithms, integration with existing network protocols & Improved explainability, enhanced adaptability & Complexity in modeling causal relationships in dynamic networks \\ 
               Large Language Model-based collective intelligence & High-performance computing infrastructure, eﬀicient model deployment  & Enhanced decision-making, improved contextual understanding & Resource intensity, potential privacy concerns \\ 
                 Generative AI integration in vehicular networks & Edge computing capabilities, low-latency communication & Real-time content generation, enhanced user experiences & Balancing model complexity with real-time requirements \\ 
                 Interactive AI with Retrieval-Augmented Generation for network management & Large-scale knowledge bases, eﬀicient retrieval mechanisms & Improved network functionality, enhanced user experience & Complexity in managing and updating knowledge bases \\

                  Generative AI agents for wireless collective intelligence & Distributed AI architectures, semantic communication protocols & Collaborative reasoning, eﬀicient knowledge transfer & Challenges in coordinating multiple AI agents \\ 

\bottomrule
\end{tabular}
\end{table*}

\section{Future Research Directions}

\subsection{Large-Scale Implementation Testing}
Large-scale implementation testing is crucial for validating the feasibility and effectiveness of AGI-enhanced IoX systems in real-world scenarios. Such testing ensures that these systems can handle the complexity, variability, and scale of practical applications, which is essential for their successful deployment and adoption. By rigorously evaluating systems in diverse and challenging environments, researchers can identify potential bottlenecks, optimize performance, and ensure reliability.

\textbf{Scalability Validation:}  
Scalability is a cornerstone of IoT systems, which must efficiently manage exponential growth in devices and data. Ensuring scalability is vital to meet the increasing demands of real-world applications without compromising performance. Research in this area focuses on:
\begin{itemize}
    \item Testing of neuro-symbolic machine learning approaches in real-world tactical edge environments.
    \item Evaluation of distributed systems performance under varying loads.
    \item Validation of multi-agent systems across diverse IoT scenarios.
    \item Testing the scalability of robust and efficient systems addressing the convergence of AI and IoT to overcome computational challenges.
\end{itemize}

\textbf{Performance Assessment:}  
Assessing performance is essential to ensure that AGI-enhanced IoX systems meet operational benchmarks in terms of speed, accuracy, and reliability. This is particularly critical in IoT environments where real-time decision-making and data processing are often required. Key directions include:
\begin{itemize}
    \item Measurement of computational efficiency in resource-constrained environments.
    \item Testing of real-time processing capabilities in complex network environments.
    \item Evaluation of system reliability and robustness under different conditions.
\end{itemize}

\subsection{Standardization of AGI Approaches}
Standardization is fundamental to ensuring interoperability, compatibility, and consistency across diverse AGI-enhanced IoX systems. By establishing common frameworks, protocols, and guidelines, standardization facilitates the seamless integration of various components and technologies, thereby accelerating development and deployment. It also promotes best practices and enables objective performance evaluation, which are critical for advancing the field.

\textbf{Architectural Standards:}  
Unified architectural standards are necessary to enable seamless communication and integration between different AI systems and components within IoX environments. This is crucial for the effective functioning of AGI-enhanced systems, which often involve multiple AI agents and devices. Research should focus on:
\begin{itemize}
    \item Development of unified frameworks for inter-AI communication.
    \item Standardization of semantic communication protocols.
    \item Creation of common interfaces for cross-layer integration.
\end{itemize}

\textbf{Implementation Guidelines:}  
Implementation guidelines ensure that AGI systems are developed and deployed consistently and effectively. Standardizing knowledge representation and evaluation metrics is particularly important for fostering collaboration and benchmarking progress. Key areas include:
\begin{itemize}
    \item Standardization of knowledge representation schemes.
    \item Development of common metrics for performance evaluation.
    \item Creation of best practices for AGI deployment in IoT environments.
    \item Standardizing blockchain-based models for crowdsensing in vehicular IoX systems to enhance privacy and scalability through decentralized technologies.
\end{itemize}

\subsection{Ethical Governance Frameworks}
Ethical governance is imperative to address the potential risks and challenges associated with AGI-enhanced IoX systems, including privacy concerns, security vulnerabilities, and accountability issues. Comprehensive frameworks ensure that these systems are developed and deployed responsibly, maintaining public trust and aligning with societal values. By prioritizing ethical considerations, researchers can mitigate adverse impacts and promote the beneficial use of AGI technologies.

\textbf{Privacy Protection:}  
Privacy protection is essential to safeguard sensitive user data and maintain trust in AGI-enhanced IoX systems. As these systems often handle vast amounts of personal information, developing robust privacy-preserving measures is critical to prevent unauthorized access and misuse. Research should focus on:
\begin{itemize}
    \item Development of privacy-preserving AGI architectures.
    \item Implementation of secure data handling protocols.
    \item Creation of user data protection frameworks.
    \item Addressing privacy concerns in IoT applications, particularly for AGI-driven systems handling sensitive data, to develop robust ethical governance frameworks.
\end{itemize}

\textbf{Accountability Mechanisms:}  
Accountability mechanisms are necessary to ensure that AGI-enhanced IoX systems operate transparently and responsibly. Establishing clear frameworks for responsibility and auditability is crucial for identifying and addressing any issues that may arise during system operation. Key directions include:
\begin{itemize}
    \item Establishment of transparency requirements.
    \item Development of auditability standards.
    \item Creation of responsibility frameworks for autonomous systems.
    \item Developing secure AGI frameworks to address security and privacy vulnerabilities of physical objects in IoT, ensuring accountability at the network and application layers.
\end{itemize}

\subsection{Quantum-Enabled Communication}
Quantum communication holds the potential to revolutionize data transmission in AGI-enhanced IoX systems by offering ultra-secure and high-speed capabilities. However, integrating quantum technologies presents unique challenges that require innovative solutions. Research in this area is essential to harness the benefits of quantum mechanics while ensuring compatibility with existing infrastructures.

\textbf{Integration Approaches:}  
Developing integration approaches for quantum communication is critical to enable its adoption in IoX systems. This includes creating network architectures that can support quantum technologies and enhancing security protocols to leverage quantum properties. Research should focus on:
\begin{itemize}
    \item Development of quantum-compatible network architectures.
    \item Investigation of quantum-enhanced security protocols.
    \item Research into quantum-resistant communication methods.
\end{itemize}

\subsection{Self-Sufficient Resource Management}
Self-sufficient resource management is vital for optimizing the use of limited resources in IoX systems, such as computational power, energy, and bandwidth. By enabling systems to autonomously manage and allocate resources, researchers can enhance efficiency and sustainability, particularly in resource-constrained or dynamic environments. This is essential for the long-term viability and scalability of AGI-enhanced IoX systems.

\textbf{Adaptive Resource Allocation:}  
Adaptive resource allocation ensures that IoX systems can dynamically respond to changing demands and conditions, optimizing performance in real-time. Intelligent and autonomous management systems are key to achieving this adaptability. Research should focus on:
\begin{itemize}
    \item Development of intelligent resource management systems.
    \item Implementation of autonomous optimization algorithms.
    \item Creation of self-organizing network architectures.
    \item Leveraging blockchain solutions for IoT to enhance secure and scalable resource management in AGI-driven IoX systems.
\end{itemize}

\textbf{Energy Efficiency:}  
Energy efficiency is crucial for IoX systems, especially those relying on battery-powered devices. Developing energy-efficient architectures and algorithms helps to prolong device lifespan and reduce environmental impact. Key directions include:
\begin{itemize}
    \item Research into energy-efficient AGI architectures.
    \item Development of sustainable computing approaches.
    \item Investigation of resource-aware algorithms.
\end{itemize}

\subsection{Advanced AI Integration Techniques}

\subsubsection{Large Language Model Integration}
Large Language Models (LLMs) offer significant potential for enhancing IoX systems through advanced natural language processing and generation capabilities. However, integrating these computationally intensive models into resource-constrained IoT devices poses significant challenges. Research in this area is essential to develop efficient deployment strategies and leverage LLMs for improved decision-making and coordination.

\begin{itemize}
    \item Developing efficient deployment strategies for LLMs on resource-constrained IoT devices, potentially through model compression or distillation techniques.
    \item Investigating Retrieval-Augmented Generation (RAG) techniques to improve decision-making by incorporating external knowledge sources.
    \item Exploring LLM-based collective intelligence systems to enable sophisticated coordination and decision-making across distributed IoX agents.
\end{itemize}

\subsubsection{Cross-Domain Knowledge Transfer}
Enhancing the ability of AGI systems to transfer knowledge across different domains is crucial for their adaptability and generalization. This is particularly important in IoX systems, where devices may need to operate in diverse and changing environments. Research should focus on developing techniques that enable effective knowledge transfer and collaborative intelligence.

\begin{itemize}
    \item Developing improved transfer learning techniques tailored for IoT devices to leverage knowledge from related tasks or domains.
    \item Investigating meta-learning approaches that enable rapid adaptation to new tasks or environments.
    \item Researching mechanisms for knowledge sharing between different AI systems to foster collaborative intelligence.
\end{itemize}

\subsubsection{Active Inference Systems}
Active inference provides a unified framework for perception, learning, and decision-making, which could enable IoX systems to operate more autonomously and efficiently. By mimicking biological processes, active inference can lead to more adaptive and intelligent systems. Research in this area is essential to develop scalable and real-time implementations suitable for large-scale IoX systems.

\begin{itemize}
    \item Developing scalable implementations of active inference that can handle the complexity of large-scale IoX systems.
    \item Creating active inference methods optimized for real-time perception in dynamic environments.
    \item Investigating the application of active inference for adaptive behavior in autonomous systems.
\end{itemize}

\subsubsection{Causal Reasoning Integration}
Integrating causal reasoning into AGI systems enhances their ability to understand cause-and-effect relationships, which is essential for informed decision-making and prediction. This is particularly relevant in IoX systems, where understanding the impact of actions on the environment is crucial. Research should focus on developing frameworks that enable effective causal inference and explanation.

\begin{itemize}
    \item Developing causal reasoning frameworks specifically designed for wireless networks to improve network management and optimization.
    \item Investigating causal inference techniques for intent management in complex systems.
    \item Researching explainable causal models that can provide insights into network behavior and facilitate optimization.
\end{itemize}

\subsubsection{Neuro-Symbolic Integration}
Neuro-symbolic approaches combine the strengths of neural networks and symbolic reasoning, offering the potential for more robust and explainable AGI systems. This integration is particularly important for IoX applications where transparency and reliability are critical. Research in this area should aim to develop methods that enhance uncertainty quantification and real-time decision-making.

\begin{itemize}
    \item Developing neuro-symbolic machine learning methods that quantify uncertainty, enhancing reliability in critical applications.
    \item Investigating explainable neuro-symbolic frameworks to improve transparency and trust in AGI systems.
    \item Researching efficient inference methods for neuro-symbolic systems to enable real-time decision-making.
\end{itemize}
\subsection{Application-Specific Research Directions}

\subsubsection{Edge Intelligence Enhancement}
Enhancing intelligence at the edge is critical for reducing latency and improving efficiency in IoX systems. By enabling real-time decision-making and data processing closer to the data source, edge intelligence can significantly enhance system responsiveness and performance. Research in this area is essential to develop advanced architectures and methods that support complex computations on edge devices.

\begin{itemize}
    \item Developing advanced preprocessing architectures that enable complex computations on edge devices.
    \item Investigating edge-based active inference methods to enable real-time decision-making without relying on centralized resources.
    \item Researching distributed cognitive architectures that allow edge devices to collaborate and share intelligence.
\end{itemize}

\subsubsection{Semantic Communications}
Semantic communications focus on transmitting meaning rather than raw data, which can significantly reduce bandwidth requirements and improve communication efficiency. This is particularly beneficial in IoX systems with limited resources. Research should explore innovative encoding schemes and protocols that enable more intelligent and context-aware communication networks.

\begin{itemize}
    \item Developing efficient semantic encoding schemes that can be standardized across different IoX systems.
    \item Investigating semantic-aware networking protocols that optimize data transmission based on content relevance.
    \item Researching semantic-level telecom world models to enable more intelligent and context-aware communication networks.
\end{itemize}

\subsubsection{Vehicular Network Intelligence}
Intelligent vehicular networks are essential for the development of smart transportation systems that can improve traffic management, safety, and efficiency. By leveraging advanced AI models and secure communication methods, these networks can enhance scene perception and decision-making capabilities. Research in this area is crucial to address the challenges of congested urban environments and ensure reliable vehicular communication.

\begin{itemize}
    \item Enhancing scene perception and decision-making capabilities through advanced AI models.
    \item Developing generative AI models tailored for vehicular networks to predict traffic patterns and optimize routing.
    \item Investigating secure spectrum access methods to ensure reliable communication in congested urban environments.
\end{itemize}

\subsubsection{Multi-Agent Systems}
Multi-agent systems enable coordinated actions across distributed IoX devices, which is crucial for handling complex tasks collaboratively. By developing frameworks and models that support intelligent behavior and collaborative reasoning, researchers can enhance the overall intelligence and efficiency of IoX systems. This is particularly relevant in scenarios where devices need to work together to achieve common goals.

\begin{itemize}
    \item Developing frameworks for coordinated multi-agent systems that can handle complex tasks collaboratively.
    \item Investigating agent-based foundation models that provide a robust basis for intelligent behavior in diverse scenarios.
    \item Researching mechanisms for collaborative reasoning among agents to enhance overall system intelligence.
\end{itemize}

\subsubsection{Air-Ground Integration}
Integrating aerial and ground-based systems can enhance coverage and flexibility in IoX networks, enabling more comprehensive and adaptable communication solutions. This integration is especially useful in dynamic environments where traditional ground-based systems may be insufficient. Research should focus on developing frameworks and protocols that seamlessly integrate air and ground networks while optimizing resource allocation.

\begin{itemize}
    \item Developing brain cognition-inspired frameworks for managing interference in integrated air-ground systems.
    \item Investigating communication protocols that seamlessly integrate air and ground networks.
    \item Researching UAV-based spectrum management techniques to optimize resource allocation in dynamic environments.
\end{itemize}
\section{Conclusion}

This survey has systematically explored the transformative potential of integrating Artificial General Intelligence (AGI) into the IoX to address critical bottlenecks in Cyber-Physical-Social-Thinking (CPST) ecosystems across sensing, network, and application layers. By synthesizing a wide range of AGI-driven architectural frameworks—spanning neuro-symbolic hybrids and active inference models at the sensing layer, semantic protocols and federated learning at the network layer, and generative AI with multi-agent reasoning at the application layer—we have demonstrated that AGI provides a robust foundation for achieving real-time adaptability, cross-domain reasoning, and context-aware decision-making. Hybrid human-artificial intelligence, as exemplified in autonomous driving, demonstrates AGI's transformative potential in real-world IoX applications by enabling advanced reasoning and decision-making in dynamic environments. These capabilities significantly surpass the limitations of traditional narrow AI, enabling IoX systems to navigate the complexities of interconnected physical, digital, social, and cognitive dimensions.
Our analysis reveals that AGI-enabled solutions effectively mitigate sensing-layer challenges, such as data overload and heterogeneous sensor fusion, through adaptive learning, selective attention, and edge preprocessing. In the network layer, joint sensing-communication-AI frameworks, blockchain-assisted federated learning, and dynamic spectrum management enhance reliability, bandwidth efficiency, and secure data handling. At the application layer, large-scale language models, structured knowledge representations, and collaborative multi-agent systems drive semantic understanding, autonomous orchestration, and scalable decision-making, with applications in smart cities, autonomous vehicles, and industrial automation.Moreover, identity modeling and addressing in IoT are crucial for managing the identity explosion in IoX application layers, with AGI-driven solutions like large language models offering significant advantages, thereby reinforcing the survey’s conclusions on AGI’s role in intelligent ecosystems. These advancements collectively position AGI-enhanced IoX as a cornerstone for next-generation intelligent systems.
Despite these strides, significant challenges persist. Balancing computational demands between edge and cloud infrastructures remains critical to address latency, energy consumption, and resource constraints. Privacy, security, and trust mechanisms must be seamlessly integrated into AGI pipelines to safeguard sensitive data and counter adversarial threats. The lack of large-scale, real-world validation for many proposed architectures raises concerns about scalability and robustness in diverse operational contexts. Moreover, the computational intensity of AGI models necessitates advancements in efficient algorithms, model compression, and hardware accelerators to ensure feasibility on resource-constrained devices.
Looking ahead, several research priorities emerge to unlock the full potential of AGI-enhanced IoX. Standardized integration frameworks and interoperable toolkits are essential to streamline AGI adoption across heterogeneous IoX platforms. Innovations in adaptive resource scheduling, lightweight inference engines, and quantum-enabled communication hold promise for overcoming computational and connectivity barriers. Ethical governance frameworks, emphasizing transparency, fairness, and accountability, must evolve alongside technical advancements to ensure responsible deployment. Rigorous field trials and cross-disciplinary collaborations will be pivotal in validating theoretical benefits and translating them into tangible operational impact.
In summary, AGI-enhanced IoX represents a paradigm-shifting frontier at the nexus of interconnected systems and advanced AI. By addressing bottlenecks across sensing, network, and application layers, AGI paves the way for adaptive, resilient, and intelligent CPST ecosystems. While challenges in scalability, efficiency, and ethical alignment remain, sustained research, standardized practices, and interdisciplinary efforts will be crucial in realizing this vision. The evolution of AGI-enhanced IoX promises to redefine the capabilities of cyber-physical-social systems, fostering a future where intelligent networks seamlessly integrate physical, digital, and cognitive interactions for societal and technological advancement.

\bibliographystyle{IEEEtran}
\bibliography{References}

\vskip 0pt plus -1fil
\begin{IEEEbiography}[{\includegraphics[width=1in,height=1.25in,clip,keepaspectratio]{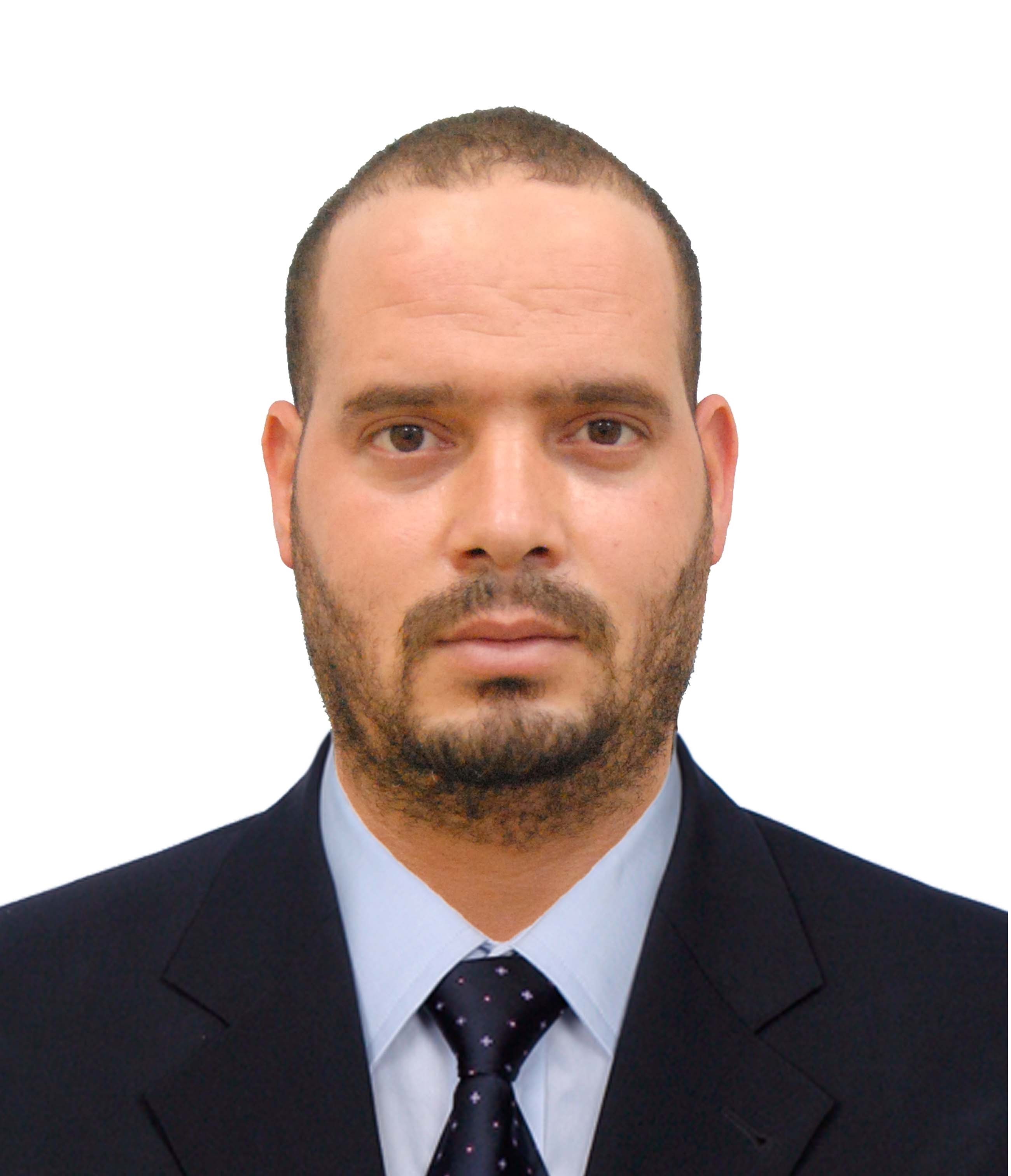}}]{Amar Khelloufi}
Received the B.S. degree (Hons.) in Computer Science from the Faculty of Sciences and Technology, Ziane Achour University of Djelfa, Djelfa, Algeria, in 2012, the M.S. degree in Distributed Information Systems from the Faculty of Sciences, University of Boumerdès, Boumerdès, Algeria, and the Ph.D. degree in Computer and Communication Engineering from the University of Science and Technology, Beijing, China, in 2024. He is currently a Research Associate at the Shenzhen Institute of Information Technology. His current research interests include the Internet of Things, Social Internet of Things, AGI, Service Recommendation, and distributed systems.

\end{IEEEbiography}
\vskip 0pt plus -1fil
\begin{IEEEbiography}[{\includegraphics[width=1in,height=1.25in,clip,keepaspectratio]{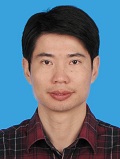}}]{Huansheng Ning}
Received his B.S. degree from Anhui University in 1996 and his Ph.D. degree from Beihang University in 2001. Now, he is a professor and vice dean of the School of Computer and Communication Engineering, University of Science and Technology Beijing, China. His current research focuses on the Internet of Things and general cyberspace. He has presided many research projects including Natural Science Foundation of China, National High Technology Research and Development Program of China (863 Project). He has published more than 200 journal/conference papers, and authored 5 books. He serves as an associate editor of IEEE Systems Journal (2013-Now), IEEE Internet of Things Journal (2014-2018), and as steering committee member of IEEE Internet of Things Journal (2016-Now).
\end{IEEEbiography}

\vskip 0pt plus -1fil

\begin{IEEEbiography}[{\includegraphics[width=1in,height=1.25in,clip,keepaspectratio]{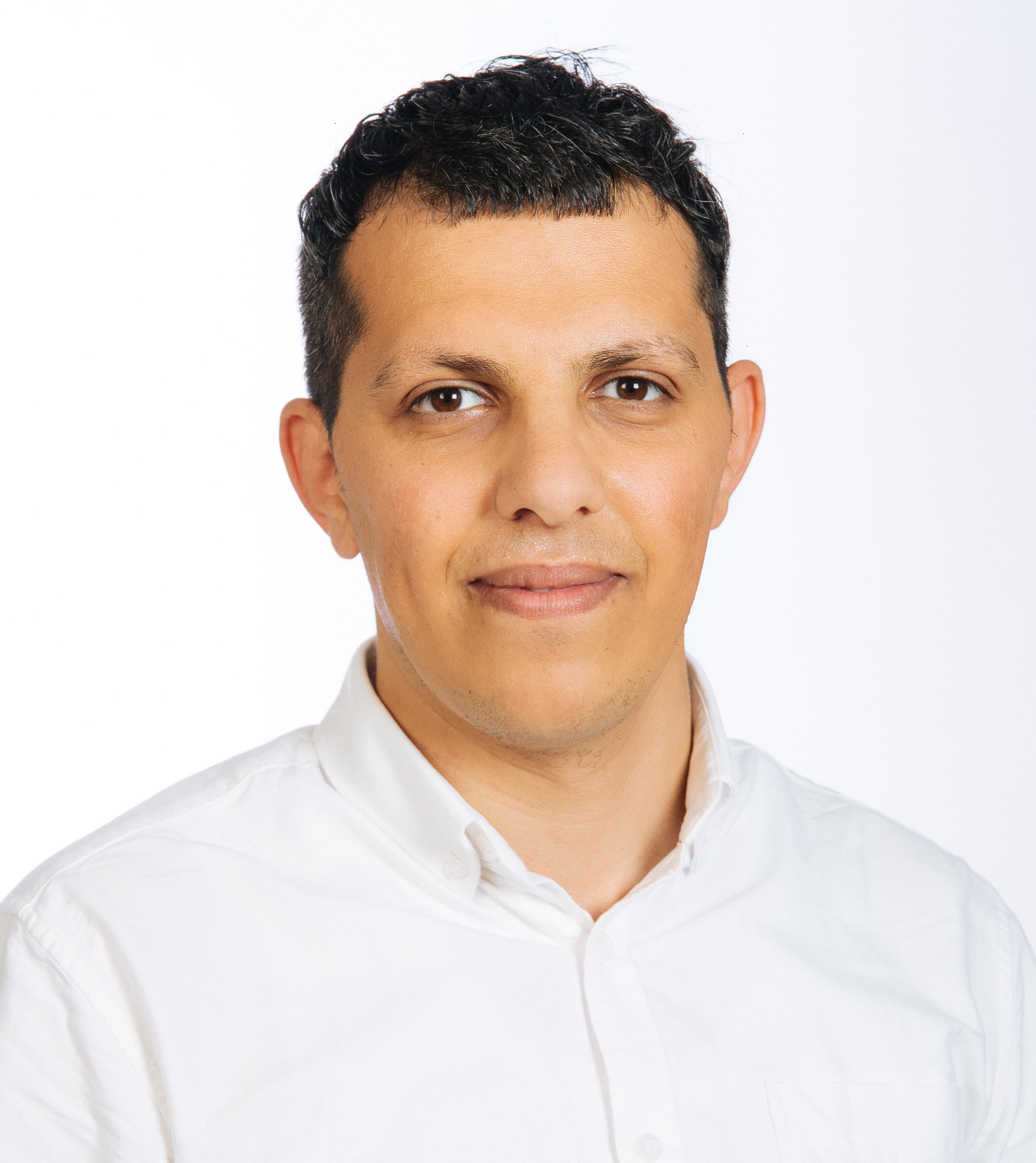}}]{Sahraoui Dhelim} is a senior researcher at University College Dublin, Ireland. He was a visiting researcher at Ulster University, UK (2020-2021). He obtained his PhD degree in Computer Science and Technology from the University of Science and Technology Beijing, China, in 2020. And a Master's degree in Networking and Distributed Systems from the University of Laghouat, Algeria, in 2014. And Bs degree in computer science from the University of Djelfa, in 2012. He serves as a guest editor in several reputable journals, including Electronics Journal and Applied Science Journal. His research interests include Social Computing, Smart Agriculture, Deep-learning, Recommendation Systems and Intelligent Transportation Systems.
\end{IEEEbiography}

\vskip 0pt plus -1fil

\begin{IEEEbiography}[{\includegraphics[width=1in,height=1.25in,clip,keepaspectratio]{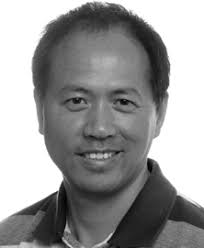}}]{Jianguo Ding } received his Doctorate in Engineering (Dr.-Ing.) from the Faculty of Mathematics and Computer Science at the University of Hagen, Germany. He is currently an Associate Professor in the Department of Computer Science at Blekinge Institute of Technology, Sweden. His research interests include cybersecurity, critical infrastructure protection, intelligent technologies, blockchain, distributed systems management and control, and serious games. He is a Senior Member of IEEE (SM'11) and a Senior Member of ACM (SM'20).

\end{IEEEbiography}

\vskip 0pt plus -1fil

\end{document}